\documentclass[a4paper,fleqn]{cas-dc}
\usepackage[numbers]{natbib}
\usepackage{color}
\usepackage{cancel}
\usepackage{ulem}
\usepackage{soul}
\usepackage[toc,page]{appendix}
\usepackage{amsfonts} 
\usepackage{graphicx}
\usepackage{subfigure}
\usepackage{color}
\usepackage{color}
\usepackage{soul}
\usepackage{cancel}
\usepackage{ulem}
\usepackage{amssymb}
\usepackage{amsmath}
\newcommand{\stkout}[1]{\ifmmode\text{\sout{\ensuremath{#1}}}\else\sout{#1}\fi}
\usepackage[switch,displaymath,mathlines]{lineno}

\begin{document}
	\let\WriteBookmarks\relax
	\def\floatpagepagefraction{1}
	\def\textpagefraction{.001}
	
	\shorttitle{Pattern formation and spatiotemporal chaos in relativistic degenerate plasmas	}
	\shortauthors{S. Das Adhikary et~al.}
	\title[mode = title]{ Pattern formation and spatiotemporal chaos in relativistic degenerate plasmas	}
	\author[1]{S. Das Adhikary}[orcid=0009-0005-3171-2507]
	\ead{sukhendusda.com}
	\author[1]{A.P. Misra}[orcid=0000-0002-6167-8136]
	\address[1]{ Department of Mathematics, Siksha Bhavana, Visva-Bharati   University, Santiniketan-731 235, West Bengal, India}
	\cormark[1]
	\ead{apmisra@visva-bharati.ac.in; apmisra@gmail.com}
	\cortext[cor1]{Corresponding author}
	\date{\today}
	%
	\begin{abstract}
		We numerically study the nonlinear interactions of high-frequency circularly polarized electromagnetic (EM) waves and low-frequency electron-acoustic (EA) density perturbations driven by the EM wave ponderomotive force in relativistic plasmas {(moderate, strong, and ultra-relativistic)} with two groups of electrons--the population of relativistic degenerate dense electrons (bulk plasma) and the sparse relativistic nondegenerate (classical) electrons, and immobile singly charged positive ions. By pattern selection, we show that many solitary patterns can be generated and drenched through modulational instability of EM waves at different spatial length scales and that the EM wave radiation spectra emanating from compact astrophysical objects may not settle into stable envelope solitons but into different incoherent states, including the emergence of temporal and spatiotemporal chaos due to collisions and fusions among the patterns with strong EA wave emission. The appearance of these states is confirmed by analyzing the Lyapunov exponent spectra, correlation function, and mutual information  {as quantitative evidence}. As a result, the redistribution of wave energy from initially exciting many solitary patterns at large scales to a few new incoherent patterns with small wavelengths in the system occurs, leading to the onset of turbulence in astrophysical plasmas. 
	\end{abstract}
	
	
	\begin{keywords}
		Pattern formation \sep Spatiotemporal chaos \sep Degenerate plasmas \sep Numerical simulation 
	\end{keywords}
	\maketitle
\section{Introduction} \label{sec-intro}
The nonthermal radiation of intense electromagnetic (EM) waves (radio wave, $X$-rays, or $\gamma$-rays) can be observed from pulsating stars and during their collapse due to compression of their magnetospheres and a significant increase in the magnetic field. The electric field in this process accelerates charged particles, resulting in charged particle radiation when they move in the magnetic field \cite{kryvdyk2009}. Thus, strong high-frequency (hf) EM waves and high-density relativistic degenerate dense plasmas associated with compact objects (e.g., white dwarfs, neutron stars, active galactic nuclei, etc.) can be simultaneously produced. Such an intense EM wave could play a crucial role in characterizing the nature of the final radiation spectra that will emerge from these compact objects. In the latter, plasmas may consist of high-density relativistic degenerate electrons, contaminated by a small fraction of classical or non-degenerate electrons and positive ions forming only the neutralizing background plasma. They thus can support the propagation of low-frequency (lf) electron-acoustic waves (EAWs) \cite{shatashvili2020nonlinear,misra2021}. Since hf EM fields can scatter into lf EM waves on EAWs, the EM spectral range will change from the original radiation and carry essential information while interacting with EAWs for the self-modulation, soliton formation \cite{shatashvili2020nonlinear,roy2022electromagnetic}, and the emergence of spatiotemporal chaos \cite{banerjee2010spatiotemporal}, leading to turbulence in astrophysical plasmas. Several authors have studied the nonlinear interactions of hf EM waves with electrostatic lf density perturbations for generating electromagnetic envelope solitons and wakefields \cite{holkundkar2018,roy2020}. However, less effort has been paid to study the emergence of spatiotemporal chaos or EM wave turbulence due to these interactions in high-density regimes.  
\par  {Typically, the fluid turbulence is associated with chaotic and multi-scale fluid flows. This occurs when fluid kinetic energy becomes high enough to overcome the dissipative effects caused by, for example, fluid viscosity, leading to a transition from large to small-scale eddies through vortex stretching in the energy cascade.
On the other hand, in the case of optical turbulence, the key physical mechanisms involve nonlinear self-focusing, which causes the high-intensity light beam to break into filaments, and modulational instability that leads to the formation of envelope solitons in dispersive media. A similar mechanism can also be applicable to plasma media.
In planetary ionospheres, plasma turbulence can occur similarly to the fluid turbulence discussed above or due to instabilities in plasmas. However, compared to solar wind or magnetospheric particles, such plasma turbulence may involve wave-particle interactions.}	
\par
Electromagnetic wave turbulence in astrophysical environments can involve interactions of several physical processes, including shocks, instabilities, and reconnection, and these interactions can result in some observable phenomena like plasma flares and jets, plasma eruptions, and sunspots. The EM wave turbulence can also be generated by other processes like linear conversion of wave modes on plasma density fluctuations, which involve resonant three-wave interactions, and fusion of electrostatic or EM waves. However, several attempts have also been made to understand wave turbulence in terms of nonlinear dynamics and chaos, which involve pattern formation and extensive and incoherent pattern dynamics in both space and time \cite{he2000pattern,he2002harmonic}. We will follow this procedure in the present investigation. In this process, the most significant conservative version of wave turbulence is the Zakharov equations (See, e.g., \cite{banerjee2010spatiotemporal}) that couple hf slowly varying electrostatic (e.g., Langmuir wave) or EM waves (linearly or circularly polarized) to slow electrostatic plasma density perturbations (e.g., electron-acoustic wave and ion-acoustic wave). Typically, decay processes (which may be an induced emission of some waves by others or nonlinear interactions of waves that result in the break-up of waves into others) accumulate wave energy into perturbations with long wavelengths. However, when this wave energy exceeds the modulational instability (MI) threshold, a transfer of energy or a redistribution of energy from initially excited large-scale modes to higher harmonic modes with small wavelengths may be possible. It is believed that some sort of chaotic process in a subsystem may be responsible for driving the energy transfer \cite{banerjee2010spatiotemporal}.  Nevertheless, even when the MI of wave envelopes is still present and the energy threshold is exceeded, the energy transfer may not occur, e.g., in the description of wave envelopes by a nonlinear Schr{\"o}inger (NLS) equation. The latter is known to be an integrable version of the Zakharov equations in the adiabatic limit (very slow-time scales) in which case the MI gets saturated in the formation of envelope solitons \cite{rizzato1998solitons}.   
\par Recently, Shatashvili \textit{et al}.  \cite{shatashvili2020nonlinear} studied the nonlinear coupling of circularly polarized EM waves and EAWs in a multi-component relativistic unmagnetized plasma with two groups of electrons (i.e., highly dense degenerate electrons and a small fraction of nondegenerate classical electrons) and immobile positive ions through the description of Zakharov-like equations.  { They predicted that even with a small fraction of non-degenerate classical electrons, electromagnetic wave envelopes can undergo modulational instability. They also obtained the instability growth rate, which is highly dependent on the degeneracy parameter. Also, they explored the existence of stationary soliton solutions in the adiabatic limit. These solutions can characterize the radiation spectra emanating from compact objects. However, they did not study in detail the influence of the degeneracy parameter $R_0$ on the profiles of the instability growth rate or the existence regimes for stable and unstable EM solitons. Their work was limited to the adiabatic limits, specifically in the solitonic region where the length scale of EM wave excitation is assumed to be small and the MI gets saturated. Beyond this scale, solitons may no longer propagate as stable structures due to the strong influence of EAW emission, potentially leading to temporal and spatiotemporal chaos. The existence of these chaotic states, which result from the redistribution of wave energy from large to small-scale perturbations, will carry essential information about the radiation spectra observed far from compact objects. This was not studied before.   }
\par 
 The purpose of this work is to extract more findings of the nonlinear coupling between the EM waves and EAWs thereby advancing the existing knowledge of EM solitons {as reported in \cite{shatashvili2020nonlinear}} by considering a region beyond the plane wave or stable solitonic region in which many linearly modes can be excited as solitary patterns and instead of settling down, saturated to new incoherent patterns after collisions and fusions due to EAW emission. We {examine the dynamics across moderate, strong, and ultra-relativistic regimes and} show that such interactions can undergo through the states of temporal chaos (TC) and spatial partial coherence (SPC), and eventually lead to the emergence of spatiotemporal chaos (STC) \cite{banerjee2010spatiotemporal}. The latter may be an essential signature of EM wave turbulence in the radiational spectra of astrophysical compact objects. {Observing these turbulent signatures is expected to provide insights into the fundamental physics associated with these extreme environments, including particle acceleration and jet formation \cite{Li2025}.  }  {Furthermore, the nonlinear and chaotic dynamics will enhance our understanding of the dynamics of degenerate relativistic plasmas and observational signatures, such as spectral broadening, EM pulse fragmentation, and temporal variability, from compact astrophysical objects.} The existence of TC and STC and the coexistence of TC and SPC are confirmed by analyzing the Lyapunov exponent spectra  \cite{rosenstein1993practical}, the two-point correlation function \cite{hohenberg1989chaotic,cai2000chaotic,he2002harmonic}, and the mutual information \cite{blahut1987principles,cai2001spatiotemporal}, {providing a dynamical route to EM wave turbulence with astrophysical implications.}
\par The manuscript is organized as follows: In Sec. \ref{sec-basic-eqs}, we present the nonlinear Zakharov-like equations, which describe the nonlinear coupling of EM waves and EAWs in multi-component degenerate plasmas and analyze the MI domains and the growth rate in weakly to ultra-relativistic degeneracy regimes. The conditions for the existence of STC are demonstrated in Sec. \ref{sec-cond-chaos}. Numerical simulation results of the coupled equations for the evolution of TC, SPC, and STC states are presented in Sec. \ref{sec-numerical}. Finally, Sec. \ref{sec-conclusion} is left to summarize and conclude the results.
\section{Basic equations}\label{sec-basic-eqs}
{One-dimensional Zakharov-like equation have been extensively studied in the context of soliton formation and emergence of chaos due to slow response of electrostatic perturbations [see, e.g., Refs. \cite{he2000pattern,he2002harmonic}]. While one-dimensional Zakharov like model has been known to be well-established model for the generation of envelope soliton through modulational instability and chaos or turbulence, two or multi-dimensional model may be useful to study wave collapse or turbulence. Since the modulational instability typically depends on the dispersion and nonlinear coefficients not the spatial coordinates, it is reasonable to consider one space dimension for such instability.   The Zakharov-like model can become quasi one-dimensional in those astrophysical conditions where EM wave turbulence is confined to a narrow, elongated structure. Such situations may arise in the environments of compact objects, where  the strong magnetic field and field-aligned electron beams restrict the wave dynamics to propagate predominantly along the magnetic field lines, effectively reducing the complex multidimensional wave dynamics to a simpler, quasi-one-dimensional Zakharov-like model.}
\par
{We aim to consider a one-dimensional model that was previously derived by Shatashvili \textit{et al}.  \cite{shatashvili2020nonlinear} in one-dimensional geometry in the context of MI of circularly polarized EM wave envelopes, and to advance the theory of considering the wave-wave interactions leading to STC. So the possibility of 2D/3D instability of solitary patterns was ruled out.}
{To start with,} we consider the nonlinear interaction of hf circularly polarized EM waves and lf EAWs in unmagnetized relativistic plasmas with two groups of electrons, namely relativistic fully degenerate dense electrons and a sparse population of relativistic nondegenerate classical electrons and stationary positive ions. The dynamics of these coupled waves is described by the following set of { one-dimensional} Zakharov-like equations \cite{shatashvili2020nonlinear}.
\begin{equation}\label{eq-EM wave}
	\begin{split}
		2i\omega_{0}\left(\frac{\partial}{\partial t}+V_{g}\frac{\partial}{\partial x}\right)A+\omega_{0} V'_{g}\frac{\partial^2 A}{\partial x^2}+\omega_{ed}^2b_{1}NA\\
		+\omega_{ed}^2b_{2}|A|^2A=0,
	\end{split}
\end{equation}
\begin{equation}\label{eq-EAW}
	\left(\frac{\partial^2}{\partial t^2}-c_{s}^2\frac{\partial^2}{\partial x^2}\right)N=-3c_{s}^2b_{3}\frac{\partial^2|A|^2}{\partial x^2},
\end{equation}
where $N\equiv\delta n_{d}/n_{d0}~(\ll1)$ stands for the degenerate electron number density perturbation normalized by the unperturbed density $n_{d0}$ such that the total number density, $n_{d}\equiv n_{d0}+\delta n_{d}$  and $A\equiv eA/mc^2$ is the slowly varying dimensionless amplitude of the EM wave envelope with $e$ denoting the elementary charge, $m$ the electron rest mass, $c$ the speed of light in vacuum, and $\omega_{0}~(k_{0})$ the EM carrier wave frequency (wave number), which satisfies the following high-frequency dispersion relation for carrier EM waves.
\begin{equation}\label{eq-hfdispersion}
	\omega_{0}^2=c^2k_{0}^2+\Omega_{d}^2+\alpha  \omega_{\rm{ed}}^2.
\end{equation}
Here, $\omega_{\rm{ed}}=\sqrt{4\pi n_{d0}e^2/m}$ is the plasma oscillation frequency of degenerate electrons,  $\Omega^2_{d}=\omega_{\rm{ed}}^2/\sqrt{1+R_{0}^2}$, $\alpha=n_{\rm{cl0}}/n_{\rm{d0}}$ ($<<1$) is the ratio of the equilibrium number densities of classical and degenerate electrons, $R_0\equiv (n_{\rm{d0}}/n_{\rm{cr}})^{1/3}=\hbar(3\pi^2n_{\rm{d0}})^{1/3}/mc$ is the electron degeneracy parameter with $n_{\rm{cr}}=m^3c^3/3\pi^2\hbar^3\approx 6\times10^{29}~\rm{cm}^{-3}$ denoting the critical number density at which the Fermi momentum, $p_F=mc$, i.e., the relativistic degeneracy effect starts. Also, $c_s$ is the electron-acoustic speed, given by, $c_s^2=\alpha c^2 R_0^2/3\sqrt{1+R_0^2}$, $V_{g}~(=d\omega_0/dk_0)$ is the group velocity of EM wave envelopes and $b_{1}=(1-\kappa^2 /\sqrt{R_{0}^2+1})$, $b_{2}=\alpha+\kappa^2/(R_{0}^2+1)^{3/2}$, and $b_{3}=(\sqrt{1+R_{0}^2}-\kappa^2)/R_{0}^2$  with $\kappa^2=1-R_{0}^2/3(1+R_{0}^2)$ pertain to the coefficients of the nonlocal nonlinear, cubic nonlinear, and ponderomotive nonlinear terms respectively. {We mention that in Ref. \cite{shatashvili2020nonlinear}, the circular polarization of EM waves was considered in the $xy$ plane and the propagation direction along the $z$-axis. However, in the model equations \eqref{eq-EM wave} and \eqref{eq-EAW}, we have considered the spatial dependence along the $x$-axis for convenience to mean that the circular polarization of EM waves is in the $yz$ plane and the propagation direction along the $x$-axis.  }
 \par { In Eq. \eqref{eq-EM wave}, $b_1$ is the coefficient of nonlocal nonlinearity. Typically, nonlocal effects significantly modify the soliton's stability and structure in contrast to the cubic or Kerr nonlinearity, which favors the formation of envelope solitons. Such nonlocal effects get enhanced due to an increase in the degeneracy parameter $R_0$. This means solitons may not be stable in strong or ultra-relativistic degenerate regimes, but can undergo chaotic states. 
 As the nonlocal nonlinearity enhances and dominates the cubic nonlinearity, the soliton becomes distorted due to collisions and fusions among the patterns by the influence of the electron-acoustic wave emission. This will eventually lead to faster redistribution of wave energy from large to small-scale solitons of higher harmonics.} 
\par From the dependency of $c_s$ on $\alpha$ (without which $c_s$ would vanish or may not be finite) it is evident that a small fraction of nondegenerate electrons is the prerequisite for the generation of low-frequency longitudinal EAWs simultaneously with the EM waves \cite{shatashvili2020nonlinear}. Equations \eqref{eq-EM wave} and \eqref{eq-EAW} were derived using a set of relativistic fluid equations for electrons with the isothermal equation of state for classical electrons and the Fermi-Dirac pressure law for degenerate species and the Maxwell's equations (See, Ref. \cite{shatashvili2020nonlinear} for details) to model the dynamics of EM wave radiation ($X$-ray and $\gamma$-ray pulses) emanating from compact astrophysical objects (e.g., white dwarf stars) and interacting with EAWs supported by degenerate dense plasmas surrounding the objects. 
\par To study numerically Eqs. \eqref{eq-EM wave} and \eqref{eq-EAW}, we normalize the space and time variables according to $t \to t\omega_{0}$ and $z \to z\omega_{0}/c_{s}$. Thus, Eqs. \eqref{eq-EM wave} and \eqref{eq-EAW} reduce to  
\begin{equation}\label{eq-nor EM wave}
	\begin{split}
		2i\left(\frac{\partial}{\partial t}+v_{g}\frac{\partial}{\partial x}\right)A+B_{0}\frac{\partial^2 A}{\partial x^2}+B_{1}NA\\
		+B_{2}|A|^2A=0,
	\end{split}
\end{equation}
\begin{equation}\label{eq-nor EAW}
	\left(\frac{\partial^2}{\partial t^2}-\frac{\partial^2}{\partial x^2}\right)N=-3b_{3}\frac{\partial^2|A|^2}{\partial x^2},
\end{equation}
where $v_{g}=V_{g}/c_{s}$, $B_{0}=\omega_{0}V'_{g}/c_{s}^2$, $B_{1}=(\omega_{\rm{ed}}/\omega_{0})^2b_{1}$, and  $B_{2}=(\omega_{\rm{ed}}/\omega_{0})^2b_{2}$.
Also, Eq. \eqref{eq-hfdispersion}, in normalized form, reduces to
\begin{equation}
	{\Omega}^2\equiv\omega_0^2/w_{ed}^2=K^2+\omega_d^2+\alpha,
\end{equation}
where $K=k_0c/\omega_{ed}$, $\Omega_d=\Omega_{d}/\omega_{ed}$ so that $v_{g}={\Omega}/K\tilde{c}_s$ with $\tilde{c}_s=c_s/c$, $B_0=(1-K^2/{\Omega}^2)/\tilde{c}_s^2$, $B_1=b_1/{\Omega}^2$, and $B_2=b_2/{\Omega}^2$. 
\par 
We inspect the spatial domains for the excitation of EM solitons through MI and their interactions with EAWs for the emergence of spatiotemporal chaos. Thus, as a first step, it is pertinent to consider the modulation of EM waves by the plane waveform of electron-acoustic density perturbations with the wave frequency $\omega$ and the wave number $k$, and to obtain the MI domain and the growth rate of instability in relativistic degenerate regimes. A plane wave solution for the EM wave envelope and the perturbed density of Eqs. \eqref{eq-nor EM wave} and \eqref{eq-nor EAW} can be written as
\begin{equation} \label{A00}
	A(x,t)=a(x,t)\exp i\theta(x,t),
\end{equation}
\begin{equation} \label{N00}
	N(x,t)=\widetilde{N}(x,t)\exp(ikx-i\omega t)+\rm{c.c.}, 
\end{equation}
where ``c.c." denotes the complex conjugate of the corresponding quantity. Next, we modulate both the amplitude and phase of the EM envelope by the plane wave perturbations as
\begin{equation}\label{eq-anstz}
	\begin{split}
		&a(x,t)=a_{0}+\tilde{a}\exp(ikx-i\omega t)+\rm{c.c.},\\ 
		&\theta(x,t)=\theta_{0}+\tilde{\theta} \exp(ikx-i\omega t)+\rm{c.c.},\\ 
	\end{split}
\end{equation}
where $a(x,t)$ and $\theta(x,t)$ are slowly varying functions of $x$ and $t$ such that $\tilde{a}<<a_{0}$, $\tilde{\theta}<<\theta_{0}$.  Substituting Eqs. \eqref{A00} and \eqref{N00} [using the ansatz  \eqref{eq-anstz}] into Eqs. \eqref{eq-nor EM wave} and \eqref{eq-nor EAW}, and looking for nonzero solutions of the amplitudes of perturbations, we obtain the following dispersion relation \cite{shatashvili2020nonlinear}
\begin{figure}[!h]
	\includegraphics[width=3.2in, height=2.3in]{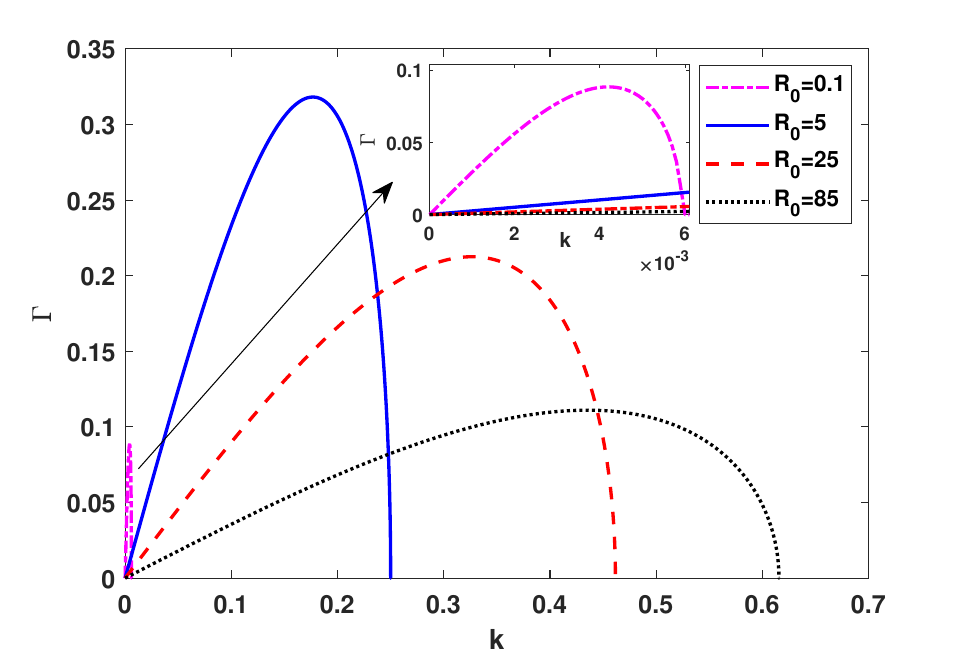}
	\caption{The modulational instability growth rate ($\Gamma$) is plotted against the modulation wave number $(k)$ for different values of the degeneracy parameter $R_0$ as in the legend. In the inset, a zoomed in part of the variation of $\Gamma$ in a small interval of $k$, especially when $R_0$ is small $(<1)$, is shown.}
	\label{fig-growth}
\end{figure}
\begin{equation}\label{eq-dispersion}
	\begin{split}
		(\omega^2-k^2)\left[(\omega-v_{g}k)^2-\frac{1}{4}B_{0}k^2(B_{0}k^2-3B_{2}a_{0}^2)\right]\\
		=\frac{3}{2}a_{0}^2B_{0}B_{1}b_{3}k^4.
	\end{split}
\end{equation}
Assuming $\omega\approx v_{g}k+i\Gamma$ and $\Gamma\ll v_{g}k$, we obtain the following growth rate of instability.
\begin{equation}\label{eq-growth}
	\Gamma\approx\sqrt{\frac{1}{4}B_{0}k^2\left[3a_{0}^2\left(B_{2}-\frac{2B_{1}b_3}{v_{g}^2-1}\right)-B_{0}k^2 \right]},
\end{equation}
provided $0<k<k_c$, where $k_c$ is the critical value of $k$ at which the growth rate $\Gamma$ vanishes, given by,
\begin{equation}
	k_c=\frac{\sqrt{3}a_{0}}{\sqrt{B_{0}}}\left(B_{2}-\frac{2B_{1}b_3}{v_{g}^2-1}\right)^{1/2}. 
\end{equation}
The maximum growth rate $\Gamma_{\max}$ occurs at $k_m=k_c/\sqrt{2}$, given by,
\begin{equation}
	\Gamma_{\max}=\frac{3}{4}a_0^2\left(B_2-\frac{2B_1b_3}{v_g^2-1}\right). 
\end{equation}
Equation \eqref{eq-growth} for $\Gamma$ agrees with Ref. \cite{shatashvili2020nonlinear} after one replaces the factors $2$ and $3$ by $3$ and $2$ respectively in the first term under the square brackets. From Fig. \ref{fig-growth}, it is seen that there are two instability subdomains of $k$: (i) $0<k\lesssim k_m$ in which the growth rate $(\Gamma)$ increases with $k$ and reaches a maximum value at $k=k_m$ and (ii) $k_m<k\lesssim k_c$ where the growth rate reduces with $k$ and vanishes at $k=k_c$. The latter thus defines a bifurcation point at which the pitchfork bifurcation occurs for unstable $(k<k_c)$ and stable $(k>k_c)$ regions.	
It is also noted that the growth rate of instability and the instability domain are significantly reduced in the regimes of weakly relativistic degeneracy (See the dash-dotted magenta line corresponding to a value of the relativistic degeneracy parameter $R_0=0.1$). However, since the degeneracy effect starts playing a role for $R_0\gtrsim1$, i.e., as one enters from moderate to ultra-relativistic degenerate regimes, the instability growth rate tends to decrease, but the instability domain expands with cut-offs at higher values of the wave number of modulation $k$ (See the solid, dashed, and dotted lines). We note that higher values of $R_0$ correspond to higher density regimes of degenerate species, and as the value of $R_0$ increases, both the group velocity dispersion $(\propto B_0)$ and the ponderomotive nonlinearity $(\propto b_3)$ tend to decrease, however, the nonlocal and the cubic nonlinearities get enhanced. As a result, the effective nonlinearity [approximately the cubic nonlinearity proportional to $B_2-3B_1b_3$ for slowly varying envelopes, which is equivalent to taking $\partial^2_tN\ll\partial^2_xN$ in Eq. \eqref{eq-nor EAW}] tends to get reduced, resulting in a reduction of the instability growth rate for the perturbed mode with a cut-off at a higher value of the wave number $k$. Typically, envelope solitons are formed due to MI, and when linearly excited modes at small scales recognize themselves into solitons, the MI gets saturated. In contrast, if the length scale of linearly excited modes is much larger than the most unstable (with maximum growth rate) ones, the EM solitons may be significantly influenced as they interact with the EAWs. Thus, in the high-density regimes $(R_0>1)$, there may be the possibilities of collisions and fusions among the modes due to EAW emission \citep{banerjee2010spatiotemporal}.  {From Fig. \ref{fig-growth}, we also note that as $R_0~(>1)$ increases, $k_m$ shifts to higher values of $k$, resulting expansions of the instability domains.   }
\section{Conditions for the existence of spatiotemporal chaos}\label{sec-cond-chaos}
Before proceeding to study Eqs. \eqref{eq-nor EM wave} and \eqref{eq-nor EAW} numerically for nonlinear EM and electrostatic wave-wave interactions, it is pertinent to state the conditions for the existence of spatiotemporal chaos in the nonlinear interactions.
\par  
A perturbation in the form of a wave $A(x,t)$ is said to exhibit spatiotemporal chaos \cite{cai2000chaotic,cai2001spatiotemporal} if the following conditions are satisfied.
\begin{itemize}
	\item $A(x,t)$ is temporally chaotic as characterized by the positive Lyapunov exponent spectra.
	\item The time series of $\{A(x_1,t)\}$ and $\{A(x_2,t)\}$ tend to become statistically independent (Spatially incoherent) as the distance between any two points $x_1$ and $x_2$ gradually increases with time $t$.
\end{itemize}
A bit more about these two conditions are as follows:
\subsection*{Temporal Chaos} We study the temporal dynamics of $A(x,t)$ and calculate the largest Lyapunov exponent spectra of the time series of $A(x,t)$ for all real values of $x$ by using the method developed by Rosenstein \cite{rosenstein1993practical}. The positive values of the largest Lyapunov exponent in the entire domain of $x$ confirm the existence of temporal chaos in the system of Eqs. \eqref{eq-nor EM wave} and \eqref{eq-nor EAW}. 
\par	
\subsection*{Statistical independence (Spatial incoherence)}
Typically, the statistical independence is estimated by the decay of two-point correlation function $C(r)$ or the decay of the mutual information function $I(r)$ with increasing values of the distance $r$ between two points. While the former is only a necessary condition, the vanishing of $I(r)$ is both a necessary and sufficient condition for the statistical independence.
We first define the correlation function as \cite{hohenberg1989chaotic,cai2000chaotic,he2002harmonic}
\begin{equation}
	C(r)\equiv C(x_1-x_2)=\left<(A(x_1,t)-\middle<A\middle>)(A(x_2,t)-\middle<A\middle>)\right>,
\end{equation}
where the angular brackets $\left<\cdot\right>$ denote the temporal mean. If $C(r)$ behaves as $\exp(-r/\xi)$ for $r\to 0$ and $\xi\ll L_x$, or $C(r)\to0$ for some $\xi$ and $r$, where $\xi $ is the correlation length and $L_x$ is the system size, then the system is said to be statistically independent or spatially incoherent. However, as in Ref. \cite{he2002harmonic}, if the correlation function lies in $0<C(r)<1$, the system may be in the spatially partial coherent state. In general, $C(r)$ lies within $-1<C(r)<1$ but may exponentially decay to zero for a particular system. 	
\par
Next, we define the mutual information function as \cite{blahut1987principles,cai2001spatiotemporal}
\begin{equation}\label{eq-mutual}
	\begin{split}
		&I(r)\equiv I(x_1,x_2)\\
		&=\int p_{x_1,x_2}(u_1,u_2)\log \left[\frac{p_{x_1,x_2}(u_1,u_2)}{p_{x_1}(u_1)p_{x_2}(u_2)}\right]du_1 du_2,
	\end{split}
\end{equation}
where the distributions $p_{x_1,x_2}(u_1,u_2)$, $p_{x_1}(u_1)$, and $p_{x_2}(u_2)$ are generated by the time series of $\{A(x_1,t)\}$ and $\{A(x_2,t)\}$ for all $t$. Intuitively, $p_{x_1,x_2}(A(x_1),A(x_2))dA(x_1)dA(x_2)$ is the fraction of time that both $A(x_1,\cdot)\in (A(x_1),A(x_1)+dA(x_1))$ and $A(x_2,\cdot)\in (A(x_2),A(x_2)+dA(x_2))$, and $p_{x_1}(A)dA$ is the fraction of time that $A(x_1,\cdot)\in(A,A+dA)$, etc. We employ the Gaussian joint probability density function to calculate the mutual information for the present system of Eqs. \eqref{eq-nor EM wave} and \eqref{eq-nor EAW}.  
{We note that while Lyapunov exponents typically quantify the exponential divergence of nearby trajectories in a chaotic system, with positive values indicating chaos and short-term predictability, correlation functions measure the relationship between a system's past and present states, exhibiting long-range order in chaotic systems and their complexity. On the other hand, mutual information assesses the statistical dependency between two variables, helping to understand information flow and complexity within a chaotic system.
 }
\section{Numerical simulation: Evolution of spatiotemporal chaos}\label{sec-numerical}
From Eq. \eqref{eq-growth} and the discussion in Sec. \ref{sec-basic-eqs}, we note that the modulation perturbation is unstable in $0<k<k_{c}$ and stable for $k>k_c$ such that $k=k_c$ defines the pitchfork bifurcation curve. The instability growth rate tends to increase in $0<k<k_{m}$ (achieving a maximum value at $k=k_m$) and decreases in $k_m<k<k_{c}$ (having a cut-off at $k=k_c$). In this regime of small growth rate, the EM waves are essentially trapped by the density humps (similar to Fig. \ref{fig-density1}), and a stationary solution of Eqs. \eqref{eq-nor EM wave} and \eqref{eq-nor EAW} in a frame $\xi=x-v_g t$ moving with the group velocity $v_g$ can be obtained as $N=3b_3|A|^2/(1-v_g^2)\sim \rm{sech}^2\left({\xi/w}\right)$ for some constant $w$. Clearly, $v_g<1$ (or the group velocity is to be smaller than the sound speed) for a soliton solution to exist.  Thus, below and sufficiently close to the bifurcation curve, i.e., in the region $k_m<k\lesssim k_{c}$, the dynamics of coupled EM waves and EAWs is truly subsonic in the sense that the MI growth rate is small $(\Gamma/k\ll1)$ \cite{rizzato1998solitons,banerjee2010spatiotemporal}. 	
So, a critical wave number $k_{c1}$ may exist in $k_{m}<k<k_{c}$ such that the motion of EM wave envelopes will be in temporal periodic or quasi-periodic states \cite{he2000pattern} if the modulation wave number lies in  $k_m<k_{c1}<k<k_{c}$, and the coexistence of temporal chaos and coherent solitary patterns may occur if $k$ lies in $k_m<k<k_{c1}<k_c$ \cite{he2000pattern}. On the other hand, the region of $k$ below $k_m$ may not be subsonic if $k_c$ is relatively large, and Eq. \eqref{eq-growth} suggests that not only is the perturbed mode with the wave number $k$ unstable but also a higher-harmonic mode with the wave number $2k$. If $k$ is further reduced from $k_m$, there is a possibility of the existence of more and more higher harmonic modes that are unstable. Thus, the region $k_m<k<k_c$ may be appropriate for describing a few unstable modes \citep{banerjee2010spatiotemporal}. We are, however, interested in the region $0<k<k_m$ where many solitary patterns may be excited through the modulational instability and saturated  after collisions, fusions, and interactions with electron-acoustic density perturbations. In the numerical simulation, we excite the solitary patterns at the modulation wavelengths $l_{m}=L_{x}/m$, where $m=1$ is for the master mode and $m=2,3,\ldots,M$ are for the unstable harmonic modes due to pattern selection such that  $M<[k^{-1}]$  \cite{he2002harmonic}.
\par We solve Eqs. \eqref{eq-nor EM wave} and \eqref{eq-nor EAW} numerically by considering a simulation box size $L_{x}=2\pi /k$ as the resonant wavelength for a perturbed mode, the number of grid points as $2048$, and the initial conditions as \cite{banerjee2010spatiotemporal,misra2009pattern,he2002harmonic}
\begin{equation}\label{eq-initial}
	\begin{split}
		A(x,0)=A_{0}[1-b\cos(kx)/L],\\
		N(x,0)=2bA_{0}\cos(kx)/L,
	\end{split}
\end{equation}
where $A_{0}$ is the amplitude of the pump EM wave envelope and $b$ is a suitable constant, and we take $L\sim500$ to mean relatively a small perturbation. Equations \eqref{eq-nor EM wave} and \eqref{eq-nor EAW} are advanced in time using the Runge-Kutta scheme with a time step $\Delta t\sim10^{-4}$ or less such that {$\Delta t< \Delta x^2\min \left\{1/B_0,1/2,1/6b_3\right\}$} (See Appendix \ref{apx-pseudocode}) is satisfied for the convergence of the scheme. We approximate the spatial derivatives by using the centered second-order difference formulas and assume {periodic boundary condition as described in appendix \ref{apx-pseudocode}}. Furthermore, we choose $A_0\sim1$ and $b\sim0.5$ and the fixed values as $K=0.01$ and $\alpha=0.03$. Thus, the key parameters are the modulation wave number $k$ and the degeneracy parameter $R_0$. While the former is responsible for the excitation of different (master and harmonic) modes in the pattern selection, the latter corresponds to the regimes of degenerate dense plasmas. Typically, the degeneracy effect starts playing a role for $R_0\gtrsim1$, i.e., in the regimes of electron number density close to or exceeding the critical number density $n_{\rm{cr}}=6\times10^{29}~\rm{cm}^{-3}$. Thus, values of $R_0$ close to $5$, $25$, and $85$ or a bit more may, respectively, be roughly called moderate, strong, and ultra-relativistic regimes. In Secs. \ref{sec-moderate}-\ref{sec-ultra}, we will mainly consider these three regimes separately. 
  Since the growth rate and the instability domain are seen to be significantly reduced for $R_0<1$ (See the dash-dotted line of Fig. \ref{fig-growth}), the case of weak relativistic degeneracy is not of interest and thus not discussed in the present study.  {We note that the MI domain for the evolution of EM wave envelopes is $0<k<k_c$ of which the subdomain $0<k\lesssim k_m$ corresponds to the region where the MI growth rate tends to increase and becomes high at $k_m$, and the excitations of large-scale EM solitary waves are possible. On the other hand, the region, $k_m<k<k_c$ corresponds to the solitonic region where the MI growth rate reduces with $k$ and gets saturated at $k=k_c$. This region exhibits the excitation of small-scale stable EM solitary waves. So, we are interested in the domain of  $0<k\lesssim k_m$. In the pattern selection, we choose three different values of $k~(<k_m)$, and we start with a value of $k$ close to but slightly lower than $k_m$, and successively reduce to choose two other values for the excitation of many solitary modes at large wavelengths. We will see that there appear mainly two subdomains of $0<k\lesssim k_m$: $0<k\lesssim k_{\rm{cs}}$ and $k_{\rm{cs}}<k\lesssim k_m$. While the former corresponds to the region for the emergence of STC, the latter is for the region where the system is in the coexistence of TC and SPC states. Here,  $k_{\rm{cs}}$ is a critical value of $k~(<k_m)$ at which a transition between these states occur. Furthermore, from Fig. \ref{fig-growth}, it is noted that the MI domain $0<k\lesssim k_m$ expands and $k_m<k<k_c$ contracts as the value of $R_0$ increases, i.e., the degeneracy regime shifts from a moderate to an ultra-relativistic regime, implying that the STC state is more likely to emerge in the ultra-relativistic degeneracy regimes with $R_0\gg1$.  }   
\subsection{Moderate relativistic degeneracy}\label{sec-moderate}
\begin{figure}[!h]
	\includegraphics[width=3in, height=1.5in]{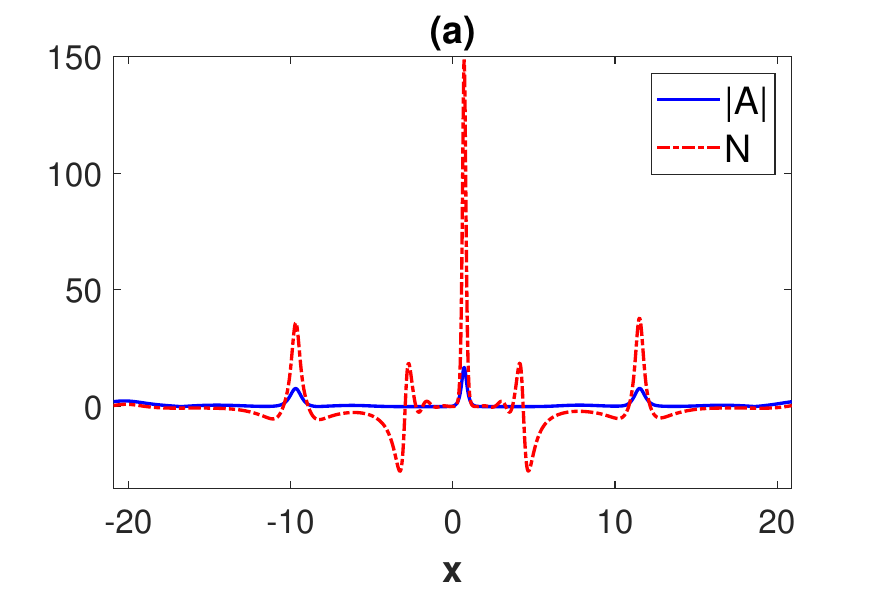}
	\includegraphics[width=3in, height=1.5in]{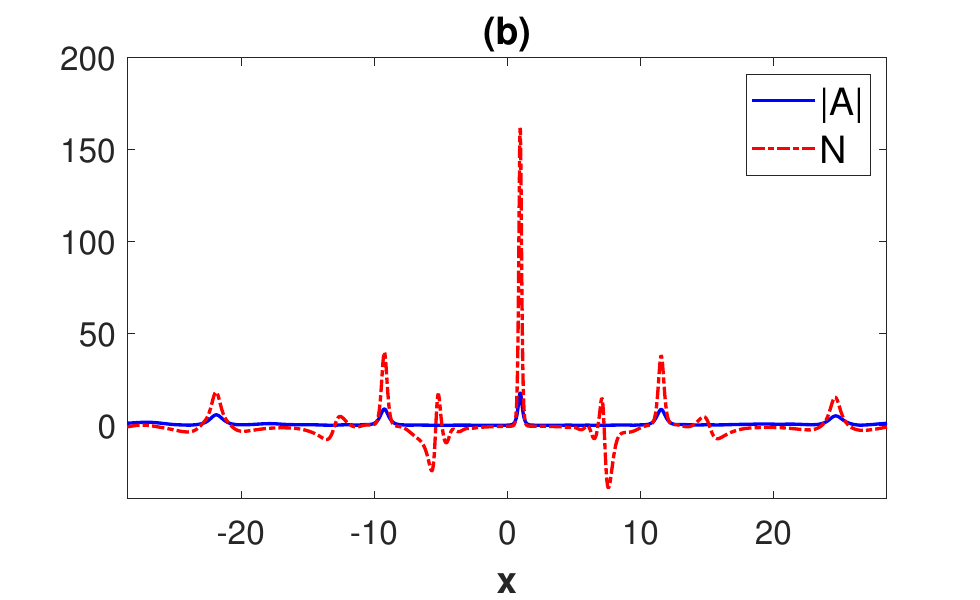}
	\includegraphics[width=3in, height=1.5in]{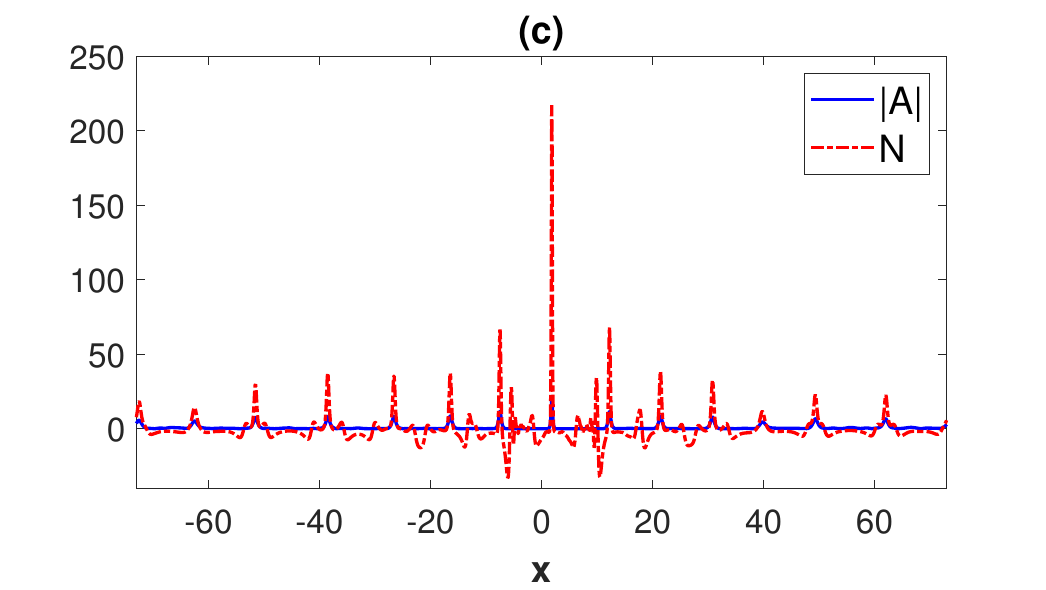}
	\caption{Initial excitations of master and harmonic solitary patterns for the EM wave field ($|A|$) and the associated density perturbation ($N$) are shown over the spatial domain in the moderate relativistic degenerate regime with $R_0=5$. It is seen that the pattern selections lead to the excitations of three, five and thirteen solitary modes corresponding to $k=0.15$ [subplot (a)],  $k=0.11$ [subplot (b)], and $k=0.043$ [subplot (c)] respectively. {The other fixed parameter values are $A_0=2$ and $b=0.5$.}}
	\label{fig-density1}
\end{figure}

We consider a relativistic degeneracy regime with $R_0\sim5$  for degenerate electrons that corresponds to the density: $n_{\rm{d0}}\sim7.5\times10^{31}~\rm{cm}^{-3}$.	{In this case, the MI domain, where the growth rate tends to become high, is  $0<k\lesssim k_m\equiv0.18$ (\textit{cf}. Fig. \ref{fig-growth}), and we choose three different values of  $k~(<k_m)$, namely $k=0.15$, $k=0.11$, and $k=0.043$ for the excitation of increasing number of wave modes ($<[k^{-1}]$)  at different wavelengths due to the pattern selection. }
Figure \ref{fig-density1} shows the spatial profiles of the perturbed EM wave field $|A|$ (solid line) and the number density perturbation of degenerate electrons $N$ associated with the EAWs (dash-dotted line) for $R_0=5$ and different values of the modulation wave number: $k=0.15$ [subplot (a)], $k=0.11$ [subplot (b)], and $k=0.043$ [subplot (c)] within the domain $0<k<k_m$ after some small interval of times, $t=12,~14$, and $30$ respectively, to observe the initial excitation of master and harmonic modes. We observe that the amplitudes of perturbations grow beyond the unity due to the modulational instability, and an excited EM field $(|A|)$ is highly correlated with the density fluctuation $(N)$ having the highest peaks near the center at $x=0$. Furthermore, in the moderate degeneracy regime, the EM waves are trapped by the density humps. Subplot (a) of Fig. \ref{fig-density1} shows that the pattern selection with $k=0.15$ leads to the excitations of three ($<[k^{-1}]=6$) modes: one master mode with wavelength $l_m=L_x$, initially peaked at $x=0$ and two harmonic modes with wavelength $l_m=L_x/4$ initially peaked at two different points: $x=-L_x/4$ and $L_x/4$ (i.e., near the points $x=-10$ and $x=10$ respectively) in the simulation box $[-L_x/2,~L_x/2]$. It is seen that the master pattern is first formed (near $x=0$) and the harmonic modes begin to appear near $x=\pm10$ from the master mode.  As we successively reduce the modulation wave number from $k=0.15$ to $k=0.11$ and $k=0.043$, the excitation of more harmonic modes (four and thirteen) with reduced wavelengths $l_m=L_x/6$ [See subplot (b)] and $l_m=L_x/14$ [See subplot (c)] are seen. The density fluctuations with $N<0$ also appear in Fig. \ref{fig-density1}, which indicates that during the pattern formation, electrons are pushed back by the EM  wave-driven ponderomotive force from the regions of the master mode and harmonic waves. Such density-depleted electrons move stochastically on either side of the patterns, and thus cannot arrest the EM wave fields \cite{banerjee2010spatiotemporal}. 
\par
\begin{figure*}
	\centering
	\includegraphics[width=6.5in, height=3in]{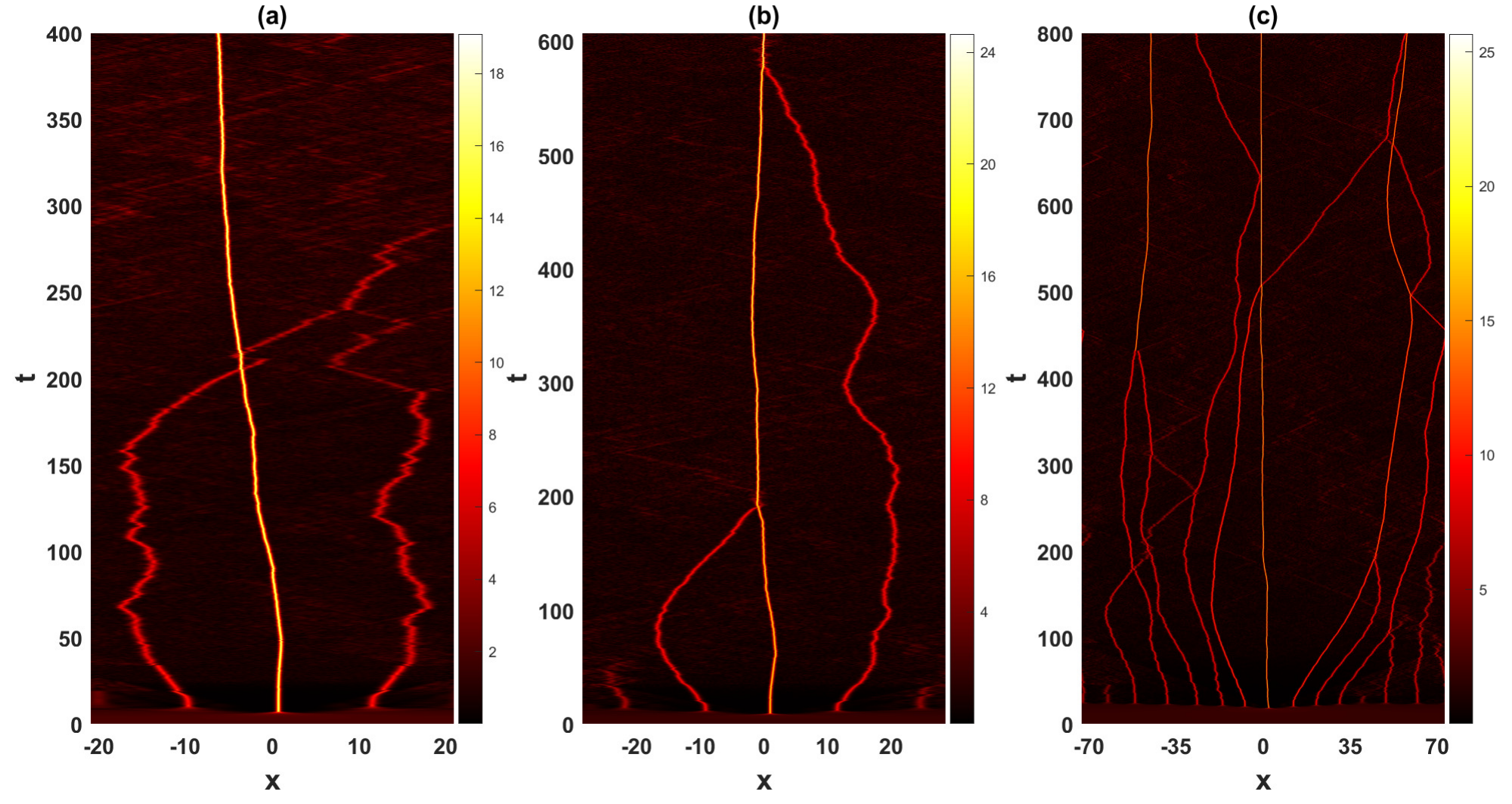}
	\caption{The EM wave field $|A(x,t)|$ is contour plotted with respect to space and time in the regime of moderate relativistic degeneracy with $R_0=5$ to show the excitation of solitary patterns (master mode and harmonic modes with different wavelengths) and their collisions and fusions into a new incoherent pattern. Subplots (a), (b) and (c) are corresponding to $k=0.15$, $k=0.11$, and $k=0.043$ respectively.}
	\label{fig-pattern1}
\end{figure*}

\begin{figure*}
	\centering
	\includegraphics[width=6.5in, height=2in]{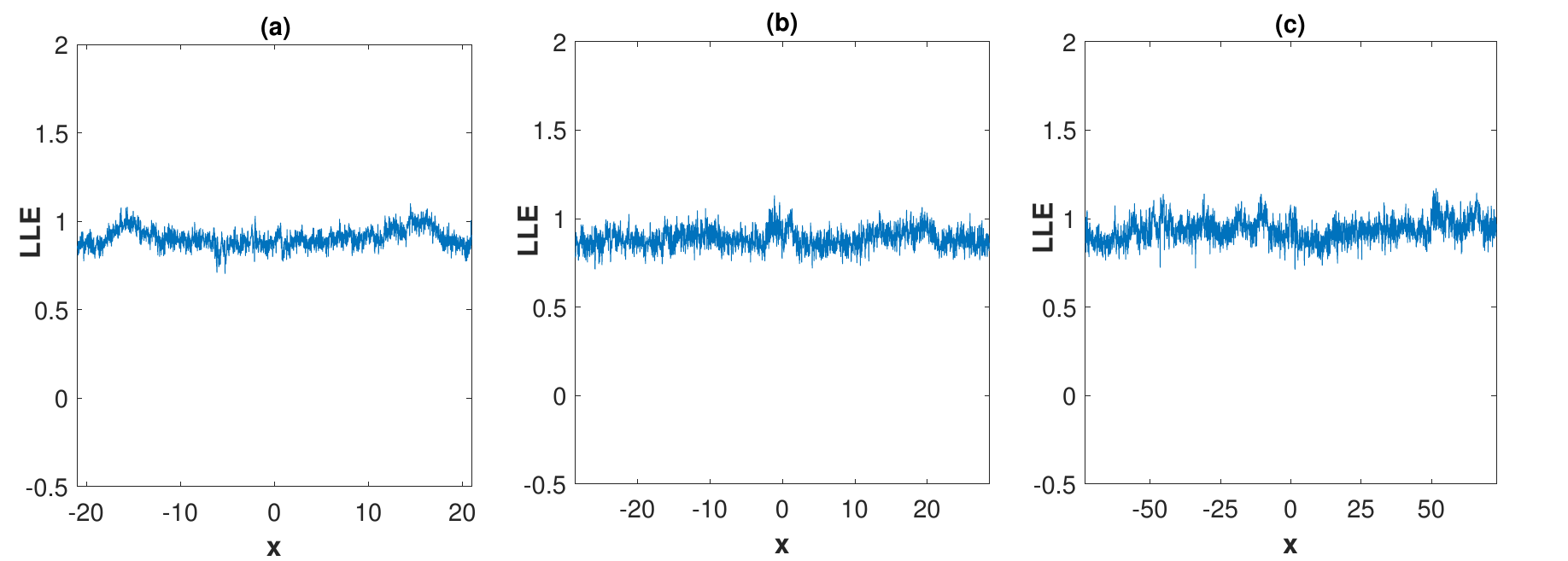}
	\caption{The largest Lyapunov exponent (LLE) spectra of the time series is shown over the spatial domain $x$ in the moderate relativistic degenerate regime with $R_0=5$ corresponding to $k=0.15$ [Subplot (a)], $k=0.11$ [Subplot (b)], and  $k=0.043$ [Subplot (c)].}
	\label{fig-lyapunov_R05}
\end{figure*}

\begin{figure*}
	\includegraphics[width=6.5in, height=2.7in]{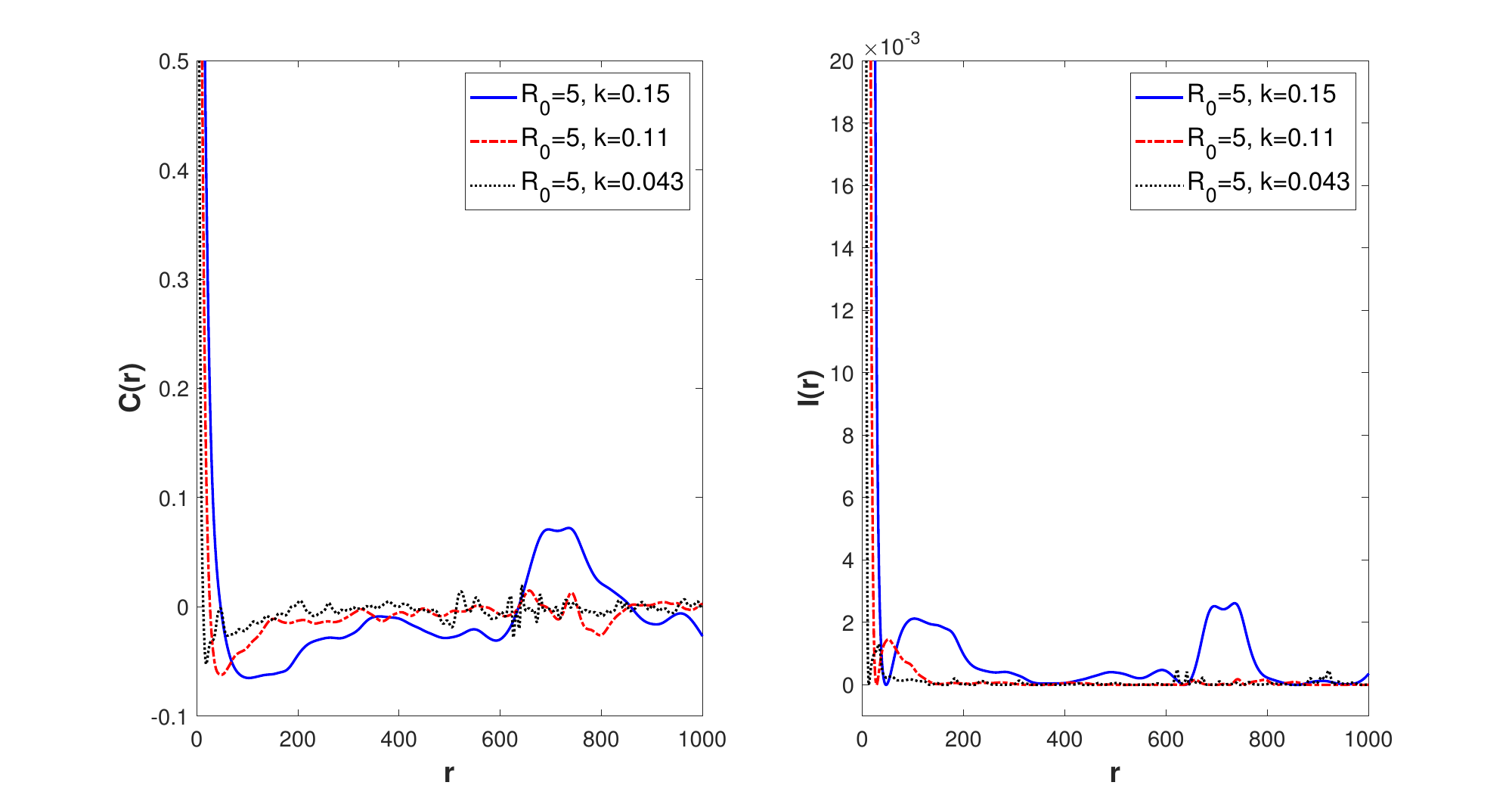}
	\caption{The correlation function [$C(r)$, subplot (a)] and the mutual information [$I(r)$, subplot (b)] are plotted against the distance $r$ between any two data points in the  moderate relativistic degenerate regime with $R_0=5$ for different values of the modulation wave number $k$ as in the legends. } 
	\label{fig-stat1}
\end{figure*}

 
As time goes on (i.e., going beyond the times $t=12,~14$, and $30$ for initial excitation of modes), some new interesting phenomena are observed. The EM master mode and harmonic modes excited at different points of $x$ with large scales collide themselves to generate higher harmonic modes with short wavelengths and also fuse after interactions due to strong EAW emission. The contour plots  (Fig. \ref{fig-pattern1}) corresponding to Fig. \ref{fig-density1}, show that for $k=0.15$ [subplot (a)] the EM master pattern (excited near $x=0$) and the two harmonic patterns (excited near $x=\pm10$) move stochastically, i.e., their amplitudes oscillate and widths vary in time, and after some time, i.e., at $t\approx 250$,  the harmonic modes collide themselves  and eventually fuse to generate a new incoherent pattern due to electron-acoustic radiation. The system energy is so weak that the new pattern does not propagate any more and eventually it disappears after $t\approx300$.  The system is then said to be in TC state as evident from the positive values of the largest Lyapunov exponent shown in the subplot (a) of Fig. \ref{fig-lyapunov_R05}. However, the spatial behaviors of the patterns are still in the SPC state due to EAW emission as evident from Fig. \ref{fig-stat1} that the correlation function $C(r)$ [subplot (a); see the solid line] still lies in the interval $-0.1<C(r)<0.1$ and the mutual information $I(r)$ [subplot (b); see the solid line] does not approach to zero as $r$ gradually increases. It follows that a few patterns may not be sufficient to cause STC, i.e., the coherence of EM waves is still partially retained and the system is in the coexistence of TC and SPC states. Subplot (b) of Fig. \ref{fig-pattern1} shows that as $k$ is further reduced from $k=0.15$ to $k=0.11$, the pattern selection leads to the excitations of five modes: four harmonic modes at $x=\pm L_x/3,~\pm L_x/6$ and a master mode at $x=0$ with initial wavelengths $L_x/6$ and $L_x$ respectively. It is observed that the patterns excited at $x=\pm L_x/3$ disappear after  $t\approx40$ due to weak EM wave energy and the influence of the strong electron-acoustic radiation. However, the other two harmonic modes initially excited at $x=\pm L_x/6$ collide at two different times $t\approx200$ and $t\approx580$ with the master mode (initially excited near $x=0$) and eventually fuse to generate a single incoherent distorted master mode with a higher amplitude but a narrower width. From subplot (b) of Fig. \ref{fig-lyapunov_R05}, the largest Lyapunov exponent is seen to be positive. Also, from subplots (a) and (b) of Fig. \ref{fig-stat1} (See the dash-dotted lines) we find an exponential decay of both the correlation function $[C(r)]$ and the mutual information $[I(r)]$. The correlation and the mutual information decay lengths, respectively, are $\xi_C \approx 9.2336$  and  $\xi_I\approx2.6781$ (See Table \ref{tab-corr-leng}), i.e., $\xi_C$, $\xi_I<<2048$, the system grid size.  So, the system is said to emerge in STC. Thus, even in the moderate relativistic regime, a few number initially excited modes can lead to the emergence of STC due to strong electron-acoustic radiation.
\par 
We also consider the case, when many unstable solitary patterns can be excited and saturated in the pattern selection, e.g., for $k=0.043$ and to verify the emergence of STC in the moderate degenerate regime. Figure \ref{fig-pattern1} (c) shows that thirteen solitary patterns can be initially formed at points $x=\pm L_x/14,~\pm l_x/7,~\pm 3L_x/14,~\pm2L_x/7,~5L_x/14,~\pm 3L_x/7$, and $x=0$ due to the pattern selection for $k=0.043$. The first collision occurs around $t=90$ between the modes excited at $x=3L_x/14$ and $5L_x/14$, and thereafter two binary collisions occur at $t\approx190$. Collisions also take place after $t=200$ in which new incoherent patterns are formed accompanied by strong electron-acoustic radiation. It is observed that three harmonic modes initially excited at $x=-3L_x/14$, $-2L_x/7$, and $-5L_x/14$ collide and fuse to generate a new higher harmonic mode with a higher amplitude but narrower width. Also, two other modes initially excited at $x=-3L_x/7$ and $2L_x/7$ disappear shortly due to influence of the electron-acoustic radiation. As time goes on, repeated collisions among the harmonic modes occur and the new incoherent patterns get distorted.  We note that the harmonic mode initially excited at $x=-L_x/14$ collides with the master mode and again collides with the new incoherent pattern formed by the repeated collisions of four solitary patterns initially excited at  $x=L_x/14,~L_x/7,~3L_x/14$,  and $-5L_x/14$. Finally, the thirteen solitary patterns with long wavelengths fuse into four incoherent patterns with small wavelengths. Form subplot (c) of Fig. \ref{fig-lyapunov_R05}, we note that the largest Lyapunov exponent is positive and from subplots (a) and (b) of Fig. \ref{fig-stat1} (dotted lines), both the correlation function and the mutual information tend to assume zero values with increasing values of $r$. The correlation and mutual information decay lengths, respectively, are $\xi_C\approx4.1894$ and $\xi_I\approx1.4613$ (See Table \ref{tab-corr-leng}), which are well below the system grid size $2048$. Thus, it may be inferred that the state of STC of the system emerges. In the STC state, the system energy, which was initially stored in thirteen solitary modes, is now spatially redistributed to four higher harmonic incoherent modes with short wavelengths in the process of random collisions and fusions among the patterns. Here, we note that a critical value $k_{\rm{cs}}$ of $k$ must exist below which a transition from TC to STC can occur. In the case of $R_0=5$, such a value lies in $0.11\lesssim k_{cs}<0.15<k_m$. Thus, in the regime of moderate relativistic degeneracy of electrons, it may be concluded that within the instability domain (where the growth rate tends to increase with $k$) but below the wave number of modulation (perturbation) $k_m$ at which the instability growth rate is maximum, the nonlinear interaction of circularly polarized EM waves and electron-acoustic density perturbations can no longer give rise to coherent solitons but lead to the emergence of spatiotemporal chaos in which energy transfer indeed occurs from initially excited many coherent harmonic modes to a few new higher harmonic incoherent modes  with short wavelengths. 
\subsection{Strong relativistic degeneracy} \label{sec-strong}
\begin{figure*}
	\centering
	\includegraphics[width=6.5in, height=3in]{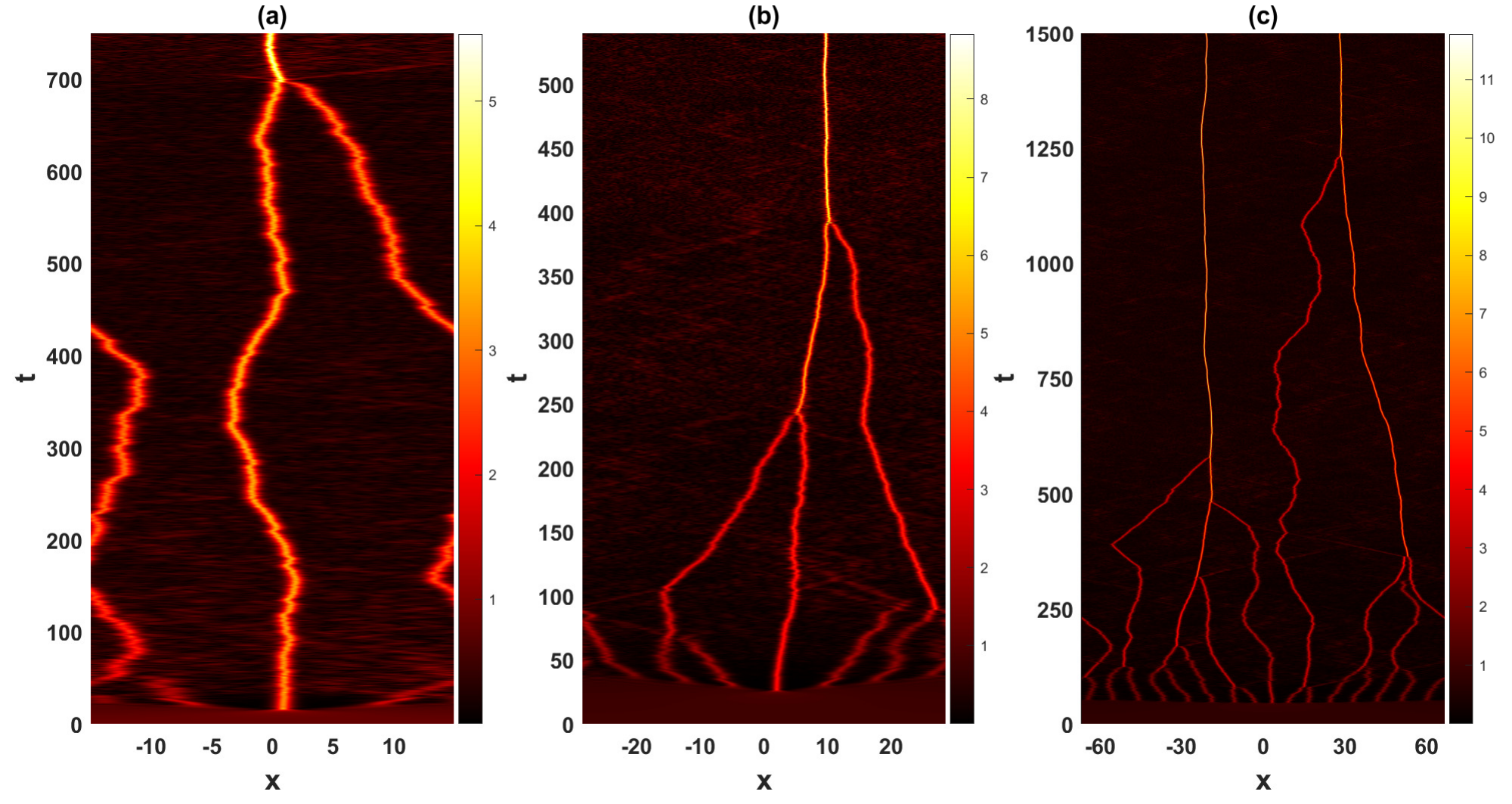}
	\caption{The EM wave field $|A(x,t)|$ is contour plotted with respect to space and time in the regime of  strong relativistic degeneracy with $R_0=25$ to show the excitation of solitary patterns (master mode and harmonic modes with different wavelengths) and their collisions and fusions into a new incoherent pattern. Subplots (a), (b) and (c) are corresponding to $k=0.21$, $k=0.11$, and $k=0.047$ respectively.}
	\label{fig-pattern2}
\end{figure*}

\begin{figure*}
	\centering
	\includegraphics[width=6.5in, height=2in]{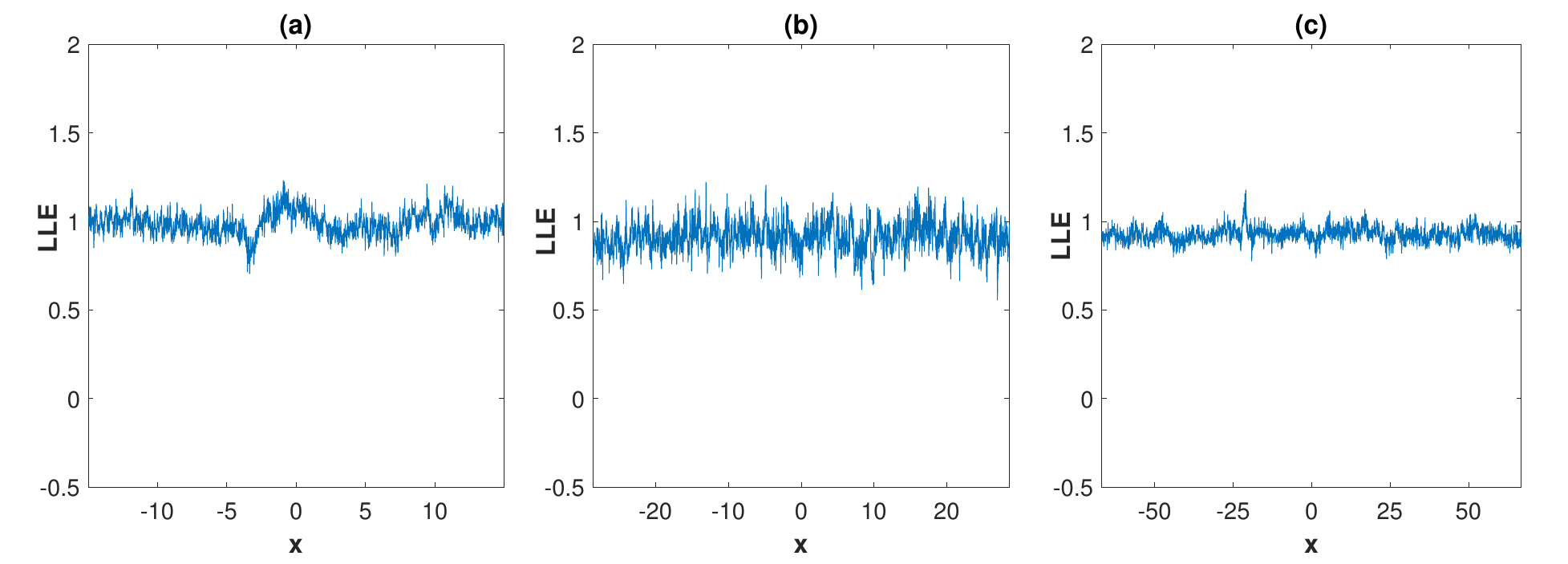}
	\caption{The largest Lyapunov exponent (LLE) spectra of the time series is shown over the spatial domain $x$ in the  strong relativistic degenerate regime with $R_0=25$ corresponding to $k=0.25$ [Subplot (a)], $k=0.11$ [Subplot (b)], and  $k=0.047$ [Subplot (c)]. }
	\label{fig-lyapunov_R025}
\end{figure*}

We consider a more dense regime in which the relativistic degeneracy of electrons becomes more stronger than the regime discussed in Sec. \ref{sec-moderate}, i.e., when $R_0=25$. This corresponds to the density regime $n_{d0}\sim 9\times10^{33}~\rm{cm}^{-3}$.
Since the MI domain expands with increasing values of $R_0$ towards higher values of the modulation wave number $k$ (\textit{cf}. Fig. \ref{fig-growth}), the TC and SPC state can coexist at a higher value of $k~(<k_m)$ than that for the case of $R_0=5$.  {In the case of $R_0=25$, the MI domain, where the growth rate tends to become high, is  $0<k\lesssim k_m\equiv0.33$ (\textit{cf}. Fig. \ref{fig-growth}), and we choose three different values of  $k~(<k_m)$, namely $k=0.21$, $k=0.11$, and $k=0.047$ for the excitation of increasing number of wave modes ($<[k^{-1}]$)  at different wavelengths due to the pattern selection. }
\begin{figure}[h!]
	\includegraphics[width=3in, height=1.5in]{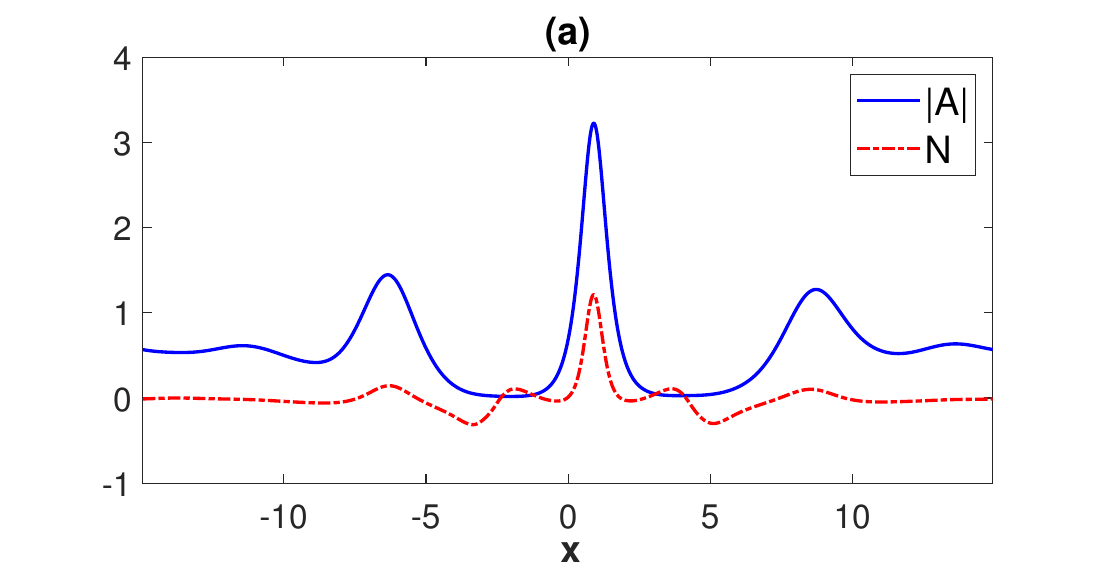}
	\includegraphics[width=3in, height=1.5in]{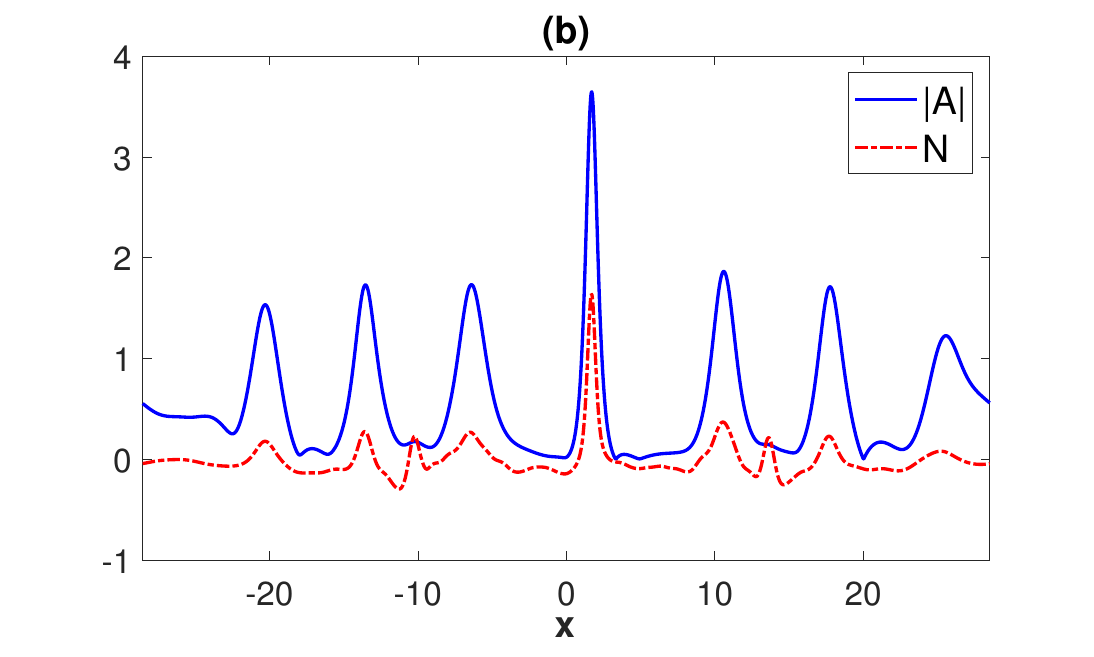}
	\includegraphics[width=3in, height=1.5in]{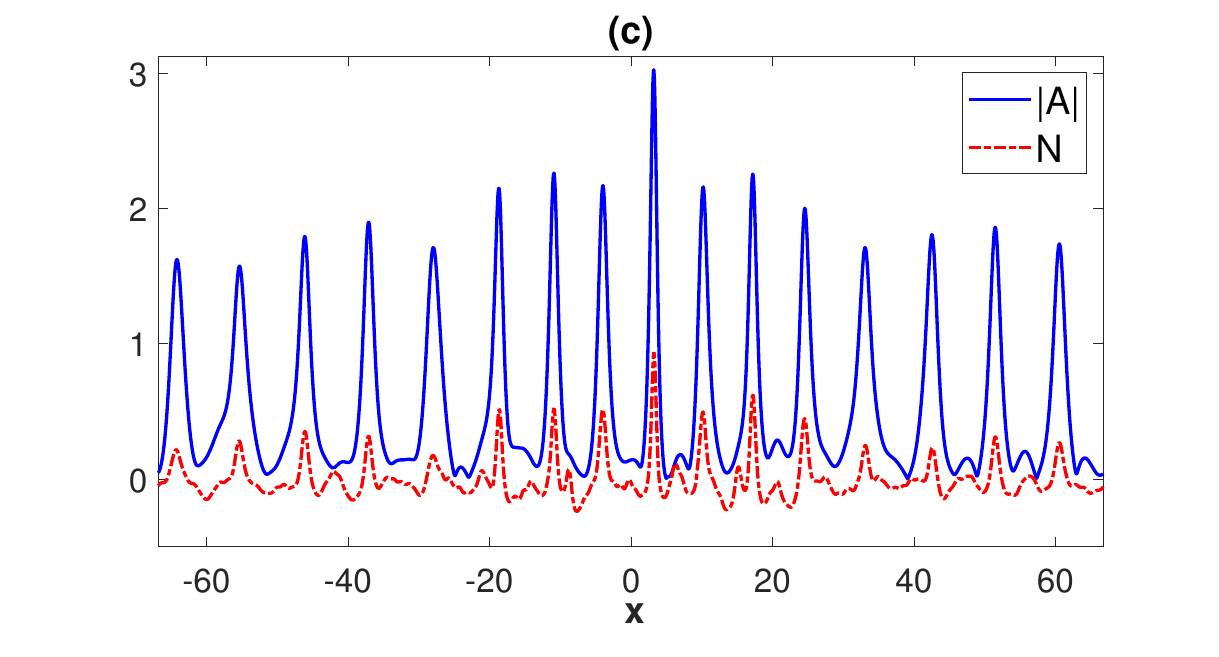}
	\caption{Initial excitations of master and harmonic solitary patterns for the EM wave field ($|A|$) and the associated density perturbation ($N$) are shown over the spatial domain in the  strong relativistic degenerate regime with $R_0=25$. It is seen that the pattern selections lead to the excitations of three, seven and sixteen solitary modes corresponding to $k=0.21$ [subplot (a)],  $k=0.11$ [subplot (b)], and $k=0.047$ [subplot (c)] respectively. {The other fixed parameter values are $A_0=0.8$ and $b=0.5$.}}
	\label{fig-density2}
\end{figure}
Figure \ref{fig-density2} shows that the pattern selections with (a) $k=0.21$, (b) $0.11$, and (c) $0.047$ initially lead to the excitations of three, seven, and sixteen solitary modes respectively. While the former (number of excited modes for $k=0.21$) remains the same, the latter two are higher than the modes excited for $R_0=5$. It follows that higher the degeneracy effects, larger are the instability domain and the excitations of modes in the pattern selection.  As before, the excited peaks of the EM patterns are correlated with the density fluctuations of EAWs. However, the amplitudes of EM waves grow stronger than the density perturbations. It implies that the electron-acoustic wave emission becomes stronger in the strong degeneracy regime. Fig. \ref{fig-pattern2}(a) shows that for $k=0.21$,  as time progresses, two of the initially excited three patterns propagate incoherently and after a long time they collide and fuse into a new incoherent pattern. The positive values of the largest Lyapunov exponent over the entire domain as in Fig. \ref{fig-lyapunov_R025} (a) manifest temporal chaos. However, from Fig. \ref{fig-stat1} (solid lines) it is seen that the correlation function and the mutual information function do not decay exponentially, and the value of the correlation function lies in $-0.2<C(r)<0.1$. So, the system is in the coexistence of TC and SPC. Although a transfer of energy from initially excited modes to a new incoherent mode occurs, it is not enough to cause STC.
\begin{figure*}
	\includegraphics[width=6.5in, height=2.5in]{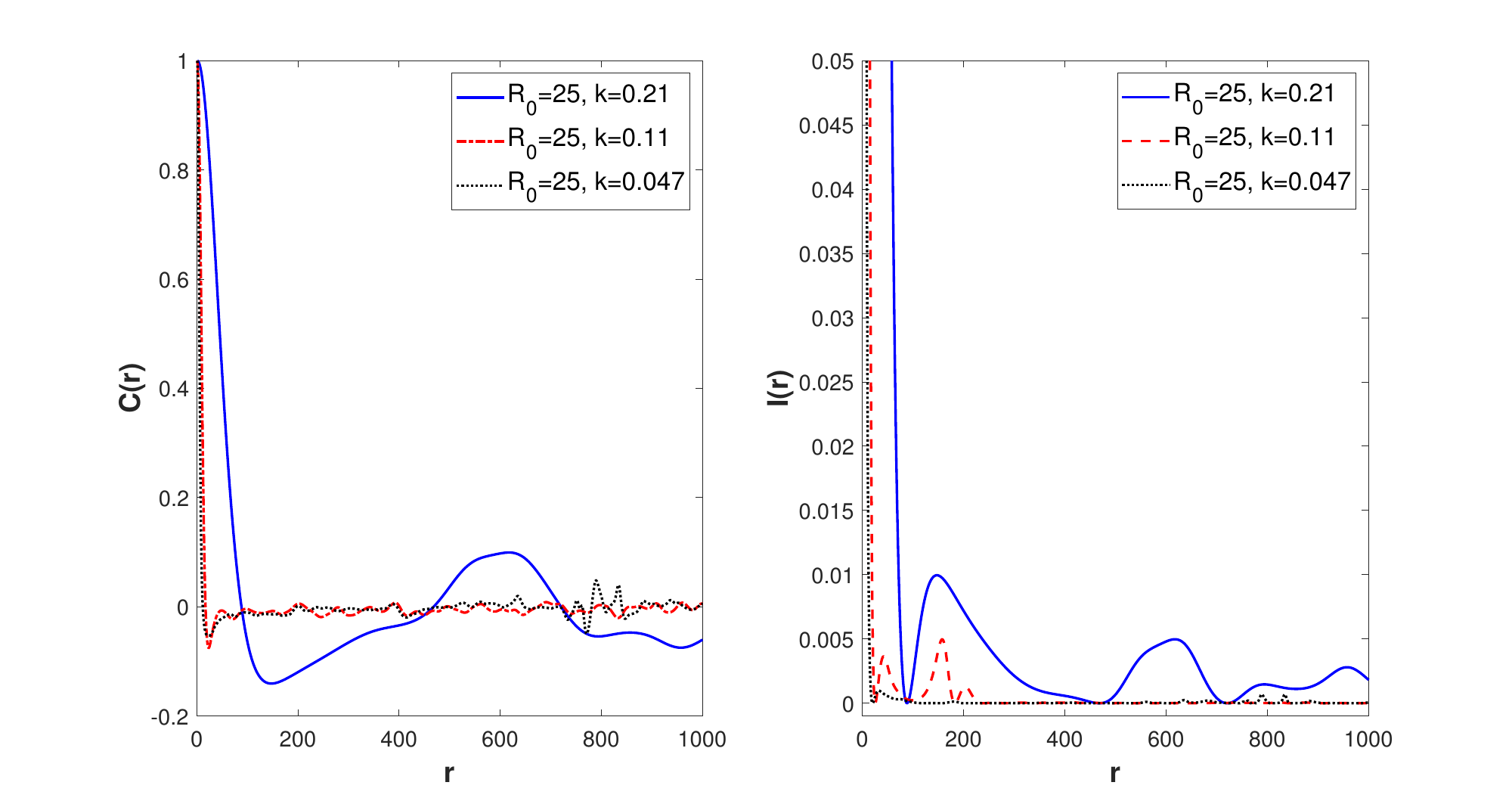}
	\caption{The correlation function [$C(r)$, subplot (a)] and the mutual information [$I(r)$, subplot (b)] are plotted against the distance $r$ between any two data points in the strong relativistic degenerate regime with $R_0=25$ for different values of the modulation wave number $k$ as in the legends.}
	\label{fig-stat2}
\end{figure*}
\par
As $k$ is reduced from $k=0.21$ to $k=0.11$, seven solitary patterns are initially excited near $x=\pm L_x/8,~\pm L_x/4,~\pm 3L_x/8$, and $x=0$  [Fig. \ref{fig-pattern2}(b)] due to pattern selection. Two of these coherent patterns, initially peaked near $x=-3L_x/8$ and $-L_x/4$, collide and fuse into a new incoherent one, and this new pattern collides with the master pattern (initially excited near $x=0$) and gets fused into another new incoherent pattern with a higher amplitude but a reduced wavelength. This newly formed pattern again collides with a pattern that is formed by the fusion of two other patterns that were initially peaked at $x=3L_x/8$ and $L_x/4$. Finally, seven coherent solitary patterns fuse into a single incoherent distorted pattern with a higher amplitude but a shorter wavelength due to electron-acoustic radiation. In this case, the largest Lyapunov exponent is seen to be positive in the entire spatial domain [Fig. \ref{fig-lyapunov_R025}(b)].  From Fig. \ref{fig-stat2} (dashed and dash-dotted lines) it is also clear that  both the correlation function and the mutual information function decay exponentially to zero [$C(r)\sim \exp(-r/7.008)$, $I(r)\sim \exp(-r/2.6846)$]. So, the system is in STC in which the spatial redistribution of energy takes place in the process of collision and fusion from initially excited seven modes to a single incoherent mode. It is noted that as the value of the degeneracy parameter $R_0$ increases, the contribution from the nonlocal nonlinearity enhances compared to a lower value of $R_0$, implying that a few number of initial patterns can also lead to STC by the influence of a strong electron-acoustic wave emission. 
\par
In order to exhibit the pattern dynamics of many initially excited modes, we further reduce the modulation wave number to $k=0.047$.  It is seen that initially excited sixteen modes [Fig. \ref{fig-density2}(c)] at different points finally fused into two incoherent patterns after several collisions and fusions among them [Fig. \ref{fig-pattern2}(c)]. The pattern initially excited at the boundary $x=-L_x/2$ does not propagate any more after $t\approx250$. However, the eight patterns initially peaked at $x=-nL_x/16$ ($n=1,...,7$) and $x=0$ fuse into one new pattern. The remaining seven patterns initially peaked at $x=nL_x/16$ ($n=1,...,7$) also collide and eventually fuse into a single new incoherent pattern. Note that in the strong degeneracy regime with reduced values of $k$, the electron-acoustic radiation also becomes stronger (since the stationary states of EM waves gradually disintegrate), which results in faster collisions and fusions among the patterns compared to the moderate degeneracy regime. The largest Lyapunov exponent spectra is seen to be positive [Fig. \ref{fig-lyapunov_R025}(c)], and the correlation function, and mutual information decay exponentially to zero [$C(r)\sim \exp(-r/3.6443)$, $I(r)\sim \exp(-r/1.5939)$; See the dotted lines of Fig. \ref{fig-stat2} for $C(r)$ and $I(r)$ and Table \ref{tab-corr-leng} for the decay lengths $\xi_C$ and $\xi_I$]. So, in this case, the system is also in STC.   For $R_0=25$, the critical wave number ($k_{cs}$) lies in $0.11<k_{\rm{cs}}<0.21$. So, STC occurs within $0<k<k_{\rm{cs}}$.

\subsection{Ultra-relativistic degeneracy with $R_0\sim85$}\label{sec-ultra}
\begin{figure*}
	\centering
	\includegraphics[width=6.5in, height=3in]{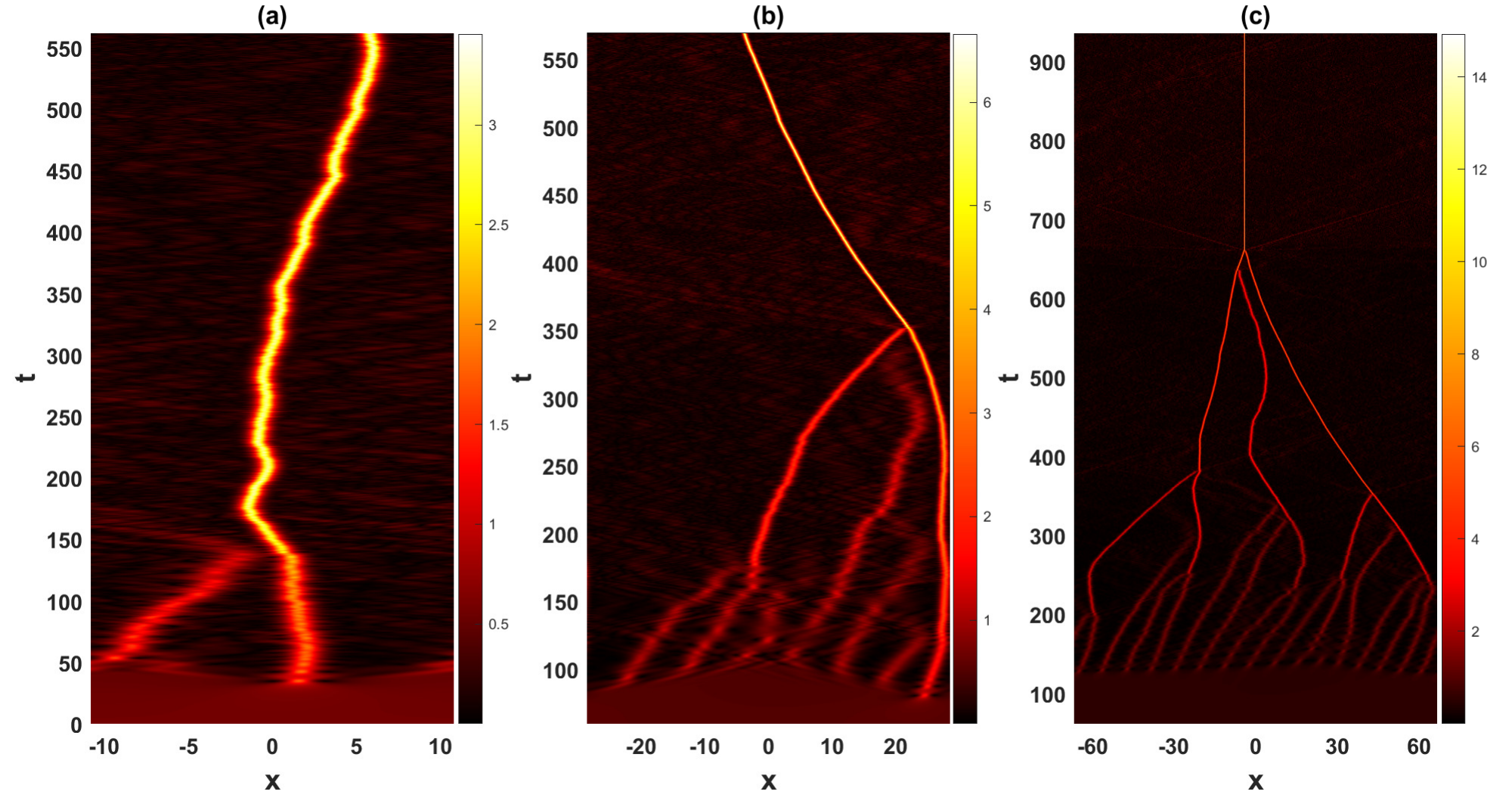}
	\caption{The EM wave field $|A(x,t)|$ is contour plotted with respect to space and time in the regime of ultra-relativistic degeneracy with $R_0=85$ to show the excitation of solitary patterns (master mode and harmonic modes with different wavelengths) and their collisions and fusions into a new incoherent pattern. Subplots (a), (b) and (c) are corresponding to $k=0.29$, $k=0.11$, and $k=0.047$ respectively.}
	\label{fig-pattern3}
\end{figure*}

\begin{figure*}
	\centering
	\includegraphics[width=6.5in, height=2in]{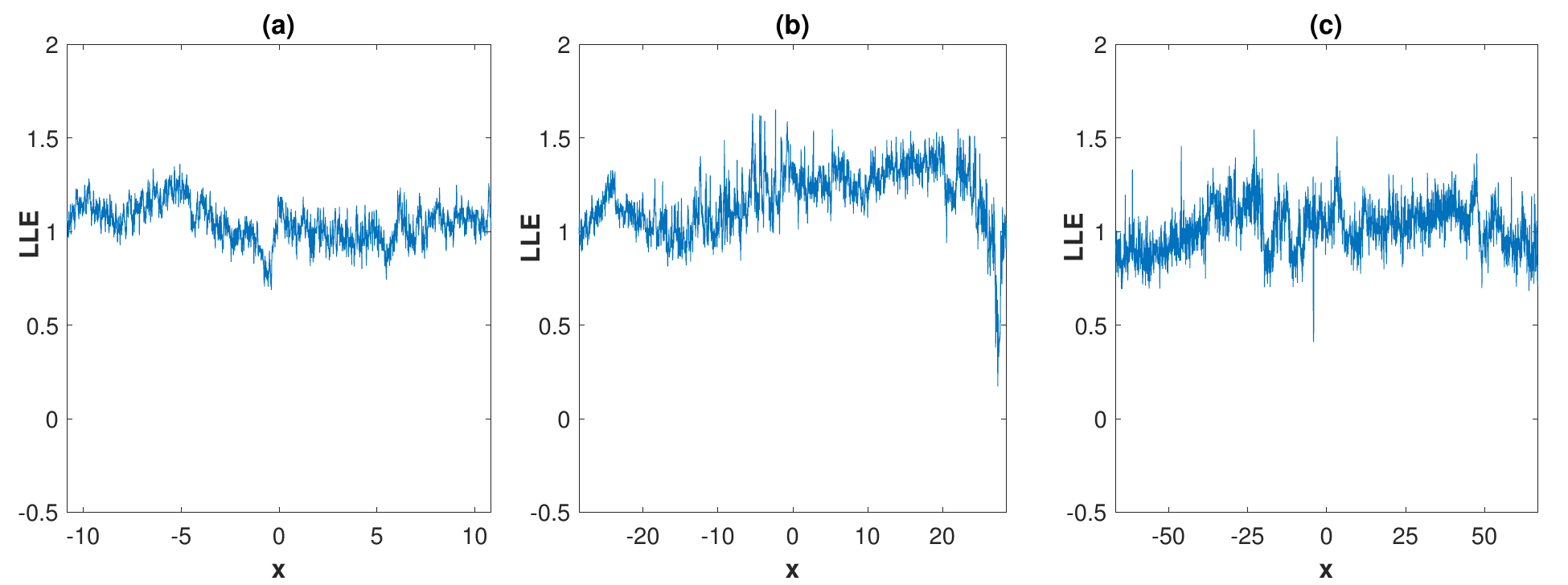}
	\caption{The largest Lyapunov exponent (LLE) spectra of the time series is shown over the spatial domain $x$ in the ultra-relativistic degenerate regime with $R_0=85$ corresponding to $k=0.29$ [Subplot (a)], $k=0.11$ [Subplot (b)], and  $k=0.047$ [Subplot (c)].}
	\label{fig-lyapunov_R085}
\end{figure*}

\begin{figure*}
	\includegraphics[width=6.5in, height=2.5in]{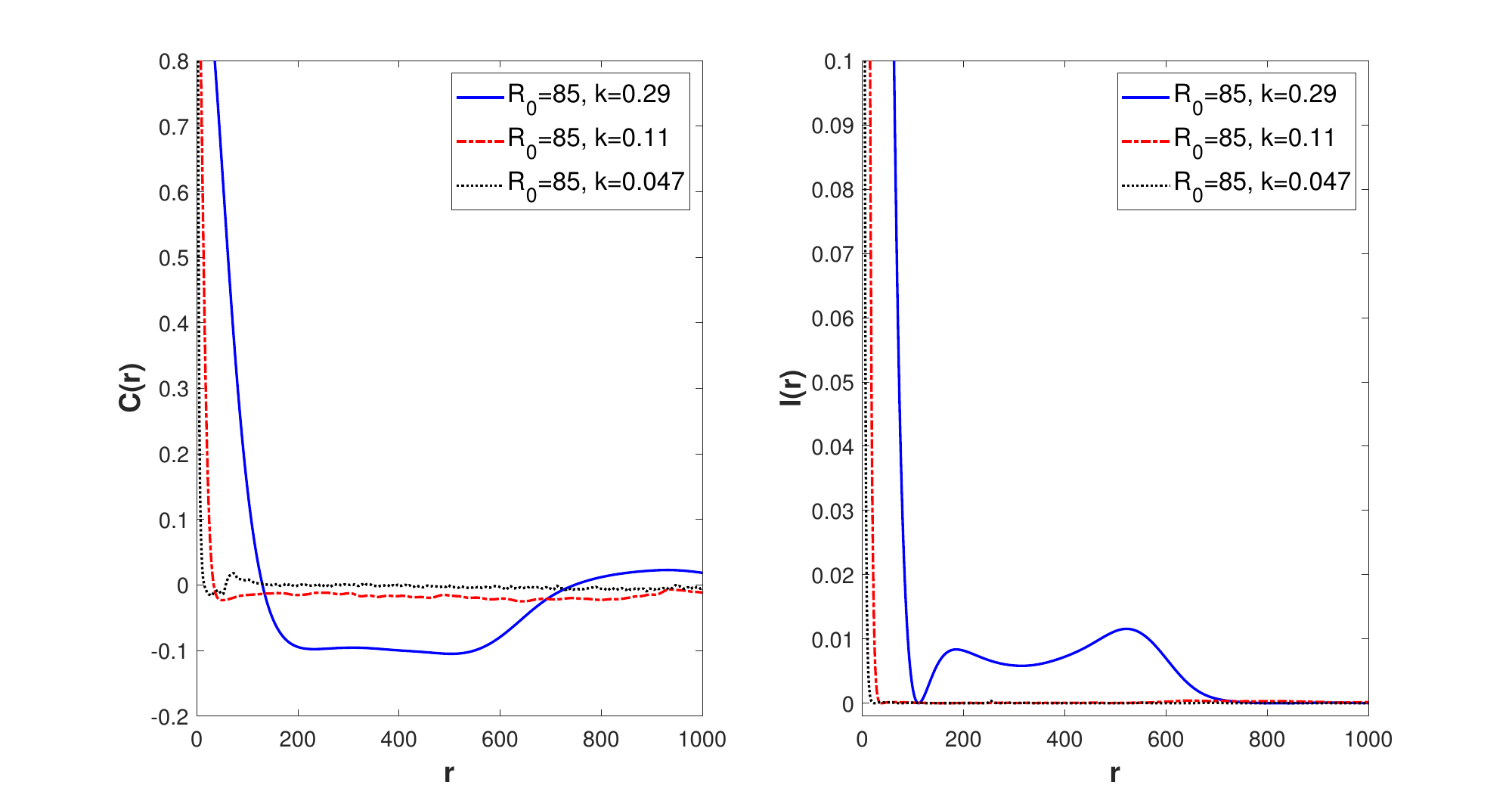}
	\caption{The correlation function [$C(r)$, subplot (a)] and the mutual information [$I(r)$, subplot (b)] are plotted against the distance $r$ between any two data points in the ultra-relativistic degenerate regime with $R_0=85$ for different values of the modulation wave number $k$ as in the legends.}
	\label{fig-stat3}
\end{figure*}

From Secs. \ref{sec-moderate} and \ref{sec-strong}, we noted that the state of STC indeed emerges in the nonlinear interaction of EM waves and EAWs in domains of the modulation wave number $(k\ll1)$ where many unstable modes can be excited initially and strong EAW emission occurs, resulting in faster collisions and fusions among the patterns, as we approach from moderate to strong relativistic degeneracy regimes with increasing values of $R_0$. It is also pertinent to examine and verify the existence of STC in the ultra-relativistic degeneracy regime in which $R_0\gg1$. For an illustration purpose, we choose $R_0=85$, which corresponds to the density regime $n_{d0}\sim 4\times10^{35}~\rm{cm}^{-3}$. However, it is free to choose any finite value of $R_0$ larger than $R_0=85$ for which the qualitative features will remain the same. In this ultra-relativistic regime, the modulational instability domain for $k$, { where the growth rate tends to become high is also higher, i.e., $0<k<k_m\equiv0.62$} (See the dotted line in Fig. \ref{fig-growth}) but with a reduced growth rate. This indicates that the system's energy transfer rate from initially excited many modes to a few modes can be faster compared to the cases of moderate and strong relativistic degeneracy regimes (See Secs. \ref{sec-moderate} and \ref{sec-strong}).
\begin{figure}[h!]
	\includegraphics[width=3in, height=1.5in]{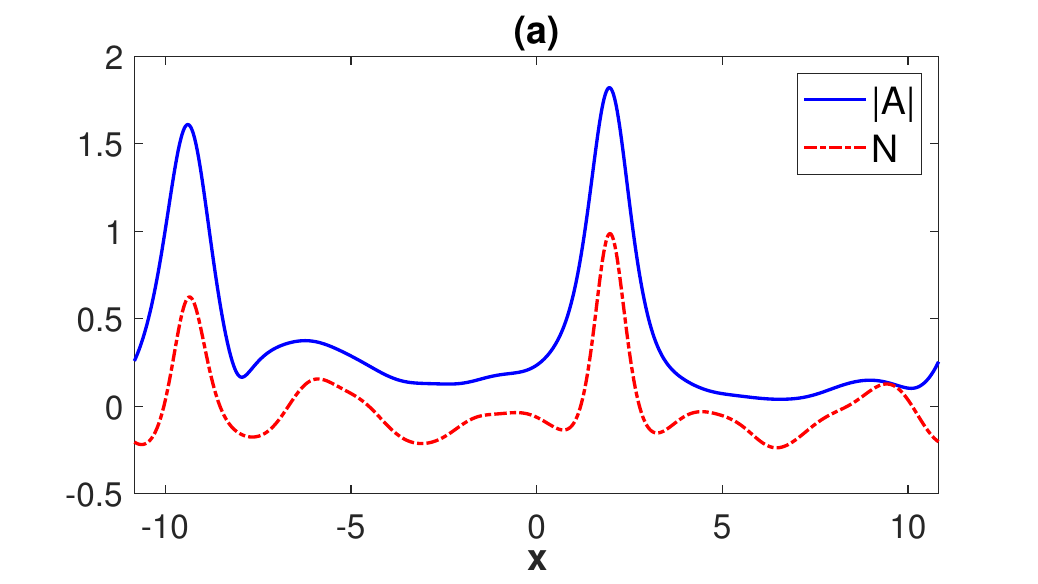}
	\includegraphics[width=3in, height=1.5in]{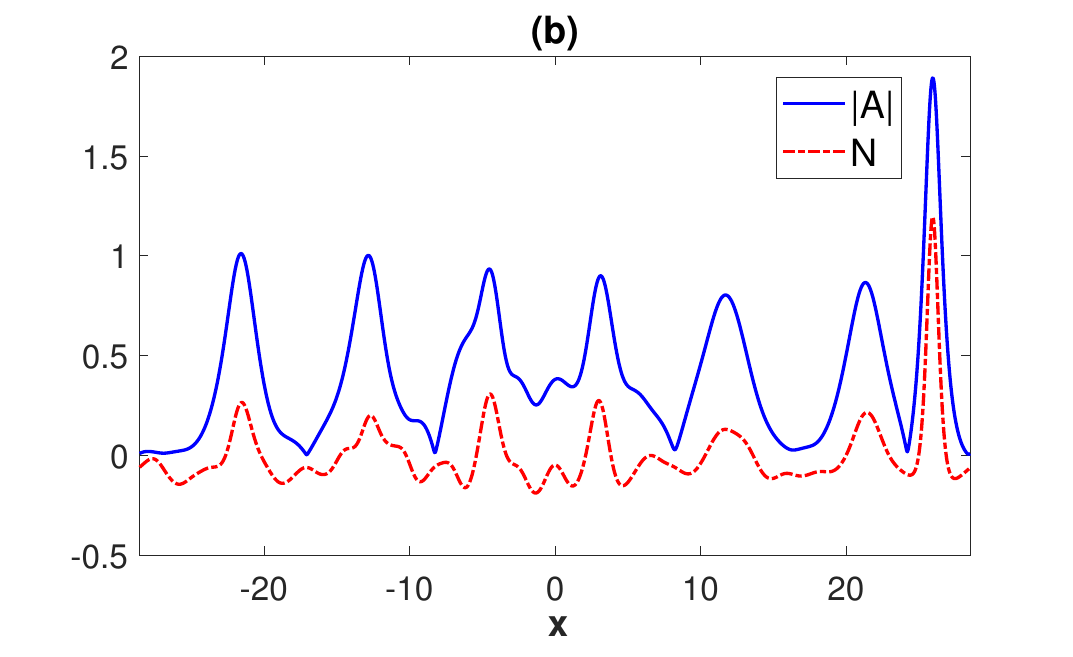}
	\includegraphics[width=3in, height=1.5in]{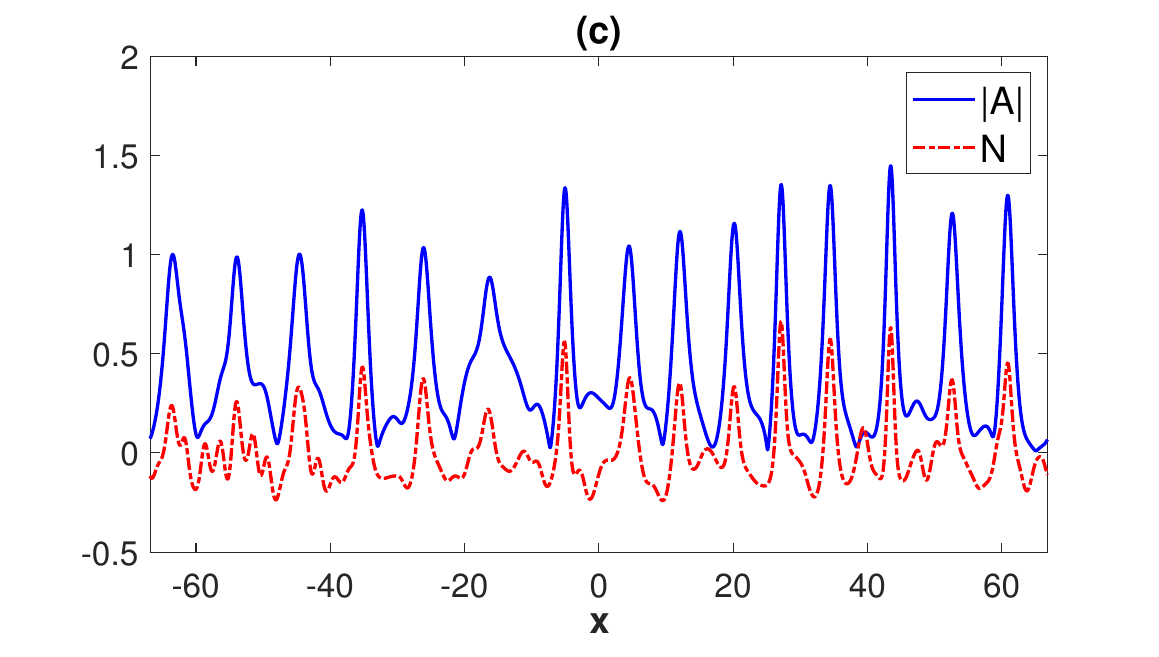}
	\caption{Initial excitations of master and harmonic solitary patterns for the EM wave field ($|A|$) and the associated density perturbation ($N$) are shown over the spatial domain in the ultra-relativistic degenerate regime with $R_0=85$. It is seen that the pattern selections lead to the excitations of two, seven and fifteen solitary modes corresponding to $k=0.29$ [subplot (a)],  $k=0.11$ [subplot (b)], and $k=0.047$ [subplot (c)] respectively. {The other fixed parameter values are $A_0=0.54$ and $b=0.4$.}}
	\label{fig-density3}
\end{figure}
As in the previous cases, we are interested in the domain $0<k<k_m\equiv0.44$. We first choose a value of $k$, which is close to but below $k_m$ and then successively reduce from it for the excitation of many modes due to pattern selection. Since modulational instability domain further expands with a cut-off of the growth rate at higher value of $k$ than that in the case with $R_0=25$, we first choose $k=0.29$ to verify the existence of TC and SPC states, { and then successively reduce it to consider the values, $k=0.11$ and $k=0.047$}. From Fig. \ref{fig-density3}, it is also evident that the EM wave peaks, initially excited at different points, are highly correlated with the density perturbations. Like Fig. \ref{fig-density2}, peaks of EMWs appear stronger in amplitudes than density perturbations, implying that the EAW emission becomes stronger in higher density regimes of degenerate electrons. For $k=0.29$, initially two solitary patterns are formed [One is the harmonic mode excited near the boundary and another master mode near $x=0$; See Fig. \ref{fig-density3}(a)]. They collide at time $t\approx140$ and fuse into a new incoherent pattern [See Fig. \ref{fig-pattern3}(a)] with a stronger amplitude but narrower width. The largest Lyapunov exponent in this case is found to be positive in the entire spatial domain of $x$ [See Fig. \ref{fig-lyapunov_R085}(a)]. However, the correlation function and the mutual information do not decay exponentially to zero [See the solid lines of subplots (a) and (b) of Fig. \ref{fig-stat3}]. Although the energy transfer takes place from initially excited two modes to a single incoherent mode, the collision and fusion are not random to cause STC. Thus, the system is in the coexistence of TC and SPC. 
\par Reducing $k$ from $k=0.29$ to $k=0.11$ gives rise to the excitation of a few more modes than two. We observe that seven patterns are formed initially due to pattern selection [See Fig. \ref{fig-density3}(b)]. Three patterns initially excited at $x=-L_x/8,~-L_x/4$, and $-3L_x/8$ collide and fuse into a new incoherent pattern. Two other pairs of patterns initially excited at $x=0,~L_x/8$ and $x=L_x/4,~3L_x/8$ collide pairwise and fuse to form two new incoherent patterns. Finally, these three new incoherent  patterns again collide at $t\approx350$ and fuse into a single new incoherent pattern with stronger amplitude but narrower width [See Fig. \ref{fig-pattern3}(b)]. The wave energy, which was initially stored in seven unstable modes, is now transferred to a single stable higher harmonic mode with a shorter wavelength \cite{he2002harmonic}. From the analysis of Lyapunov exponent spectra and statistical analysis, it is noted that the largest Lyapunov exponent is positive [See Fig. \ref{fig-lyapunov_R085}(b)], and the correlation function and mutual information decay exponentially to zero [$C(r)\sim \exp(-r/12.2684)$, $I(r)\sim \exp(-r/3.9002)$; See the dash-dotted lines of subplots (a) and (b) of Fig. \ref{fig-stat3} and Table \ref{tab-corr-leng}]. Thus, in the ultra-relativistic degeneracy regime, not only occur the random collisions and fusions among the solitary patterns within a shorter interval, the final energy transfer from seven patterns to a new incoherent pattern also becomes faster than the case with $R_0=25$. Evidently, the system is in STC. 
\par In order to illustrate more frequent collisions and fusions among patterns and faster energy transfer at a reduced modulation wave number $k$ compared to the previous case, we consider $k=0.047$. In the latter, fifteen patterns are seen to be excited initially due to pattern selection [See Fig. \ref{fig-density3}(c)]. From  \ref{fig-pattern3}(c), we note that the stochastic motions of pattern trains lead to frequent collisions among the neighboring patterns and fuse into new incoherent patterns. Finally, initially excited fifteen patterns fuse into a single incoherent pattern with a higher amplitude but a narrower width at a time shorter than the case with $R_0=25$ [See Fig. \ref{fig-pattern2}(c)]. From the positive values of the largest Lyapunov exponent over the entire spatial domain [Fig. \ref{fig-lyapunov_R085}(c)] and the decays of the correlation function and the mutual information (with $\xi_C\approx 4.3995$, $\xi_I\approx1.2533$; See Table \ref{tab-corr-leng} and the dotted lines of Fig. \ref{fig-stat3}) it is evident that the STC state also emerges in the ultra-relativistic regime. In this case, the critical wave number ($k_{\rm{cs}}$) lies in $0.11<k_{\rm{cs}}<0.29$.  
\begin{table*}[!h]
	{
		\centering
		\renewcommand{\arraystretch}{1.6}
		\begin{tabular}{|p{3cm}|p{3cm}|p{2.5cm}|p{1.7cm}|p{2.5cm}|}
			\hline
			\textbf{Relativistic Degeneracy ($R_0$)}& \textbf{MI domain ($\Gamma$ tends to become high) Large-scale excitation ($0<k\lesssim k_m$) } & \textbf{Pattern selection $\left(M=\left[k^{-1}\right]\right)$; Excitation of increasing number of EM solitary waves at different $k<k_m$}&\textbf{Appearance of different states} & \textbf{ Critical wave number ($k_{cs}$) at which transition from one state to another occurs}\\
			\hline
			& & \smash{\(k = 0.15\)}    &TC \& SPC&  \\
			\cline{3-4}
			& & \smash{\(k = 0.11\)}   &STC& \\
			\cline{3-4}
			\multirow{-3}{3cm}{Moderate: $R_0=5$, i.e., $n_{\rm{d0}}\sim8\times10^{31}~\rm{cm}^{-3}$} &\multirow{-3}{3cm}{$0<k\lesssim k_m\equiv0.18$ }  & \smash{\(k = 0.043\)} &\small STC &\multirow{-3}{2.5cm}{$k_{cs}\in\left(0.11,0.15\right)$ } \\
			\hline
			& & \smash{\(k = 0.21\)}    &TC \& SPC&  \\
			\cline{3-4}
			& & \smash{\(k = 0.11\)}   &STC& \\
			\cline{3-4}
			\multirow{-3}{3cm}{Strong: $R_0=25$, i.e., $n_{\rm{d0}}\sim9\times10^{33}~\rm{cm}^{-3}$} &\multirow{-3}{3cm}{\smash{$0<k\lesssim k_m\equiv0.33$} }  & \smash{\(k = 0.047\)} &\small STC &\multirow{-3}{2.5cm}{$k_{cs}\in\left(0.11,0.21\right)$ } \\
			\hline
			& & \smash{\(k = 0.29\)}    &TC \& SPC&  \\
			\cline{3-4}
			& & \smash{\(k = 0.11\)}   &STC& \\
			\cline{3-4}
			\multirow{-3}{3cm}{Ultra-relativistic: $R_0=85$, i.e., $n_{\rm{d0}}\sim4\times10^{35}~\rm{cm}^{-3}$} &\multirow{-3}{3cm}{$0<k\lesssim k_m\equiv0.44$ } & \smash{\(k = 0.047\)} &\small STC &\multirow{-3}{2.5cm}{$k_{cs}\in\left(0.11,0.29\right)$ } \\
			\hline
		\end{tabular}
	}
	\caption{A comparative summary showing the emergence of different states of the system in three different regimes of the relativistic degeneracy ($R_0$).}
	\label{tab-summary}
\end{table*}

\begin{table}[!h]
	\centering
	\begin{tabular}{|l|l|l|l|l|l|l|}
		\hline
		& \multicolumn{2}{c|} {$R_0=5$}& \multicolumn{2}{c|}{$R_0=25$}&\multicolumn{2}{c|}{$R_0=85$}\\
		\hline
		$k$&$0.11$&$0.043$&$0.11$&$0.047$&$0.11$&$0.047$\\
		\hline
		$\xi_C$&$9.2336$&$4.1894$&$7.0028$&$3.6443$&$12.2684$&$4.3995$\\
		\hline
		$\xi_I$&$2.6781$&$1.4613$&$2.6846$&$1.5939$&$3.9002$&$1.2533$\\
		\hline
	\end{tabular}
	
	\caption{The values of the correlation decay length ($\xi_C$) and the mutual information decay length ($\xi_I$) for $C(r)\sim \exp(-r/\xi_C)$ and $I(r)\sim\exp(-r/\xi_I)$. Clearly, $\xi_C$, $\xi_I<<2048$. So, for these cases, the system exhibits STC.}
	\label{tab-corr-leng}
\end{table}
\par 
{To compare the results in Secs. \ref{sec-moderate}--\ref{sec-ultra} across three degeneracy regimes, we present a summary in Table \ref{tab-summary}. As the degeneracy regime changes from moderate to strong to ultra-relativistic with increasing $R_0$, the intervals admitting the critical value $k_{\rm{cs}}$ expand. As a result, the domains of $k$ for both the coexistence of TC and SPC states, as well as the presence of STC states, also increase. Thus, the STC state appears in a wider domain of $k$ under ultra-relativistic degeneracy than under moderate or strong degeneracy. Since the MI domain grows with increasing $R_0$, shifting from moderate to ultra-relativistic regimes reduces the possibility of stable EM solitons but enhances the emergence of STC states.}
\section{Discussion and conclusion}\label{sec-conclusion}	
We have studied the nonlinear interactions of circularly polarized EM waves and electron-acoustic density perturbations in a relativistic degenerate unmagnetized plasma with two groups of electrons, namely, classical (sparsely populated) and degenerate (densely populated) electrons, and stationary singly charged positive ions. We show that the interactions can lead to the formation of EM envelope solitons via the modulational instability of plane wave perturbations. The instability growth rate tends to get reduced significantly and the instability domain for the modulation wave number expands as one approaches from weakly relativistic to ultra-relativistic degeneracy regimes. By a numerical simulation approach, we show that by the pattern selection, many solitary patterns can be initially formed by the excitations of harmonic modes from the spatially modulated master mode. If the length scale of excited modes is smaller than the most unstable one (having maximum growth rate at $k_m$), the EM solitons can propagate as stable modes in the instability domain $k_m<k<k_c$. However, in contrast, if the length scale of excitations of harmonic modes is larger than the most unstable mode (i.e., in the instability domain $0<k<k_m$), the EM solitons may be significantly influenced as they interact with the electron-acoustic perturbations and the collisions and fusions among the modes may likely occur due to EAW emission. In this situation, the interactions can be either in the states of TC and SPC, or in the STC state, depending on how much the length scale of excitation is larger than the most unstable mode and whether the collisions and fusions among the patterns are random or stochastic in nature.
\par 
To elucidate the characteristics of the wave-wave interactions in degenerate plasmas, we consider three different values of the degeneracy parameter: $R_0=5,~25$, and $85$ to characterize moderate, strong and ultra-relativistic degeneracy regimes with degenerate electron number densities  $n_{\rm{d0}}\sim8\times10^{31}~\rm{cm}^{-3}$, $9\times10^{33}~\rm{cm}^{-3}$, and $4\times10^{35}~\rm{cm}^{-3}$ respectively.    In all these cases, the pattern selection restricts the number of excited harmonic modes to be smaller than $[k^{-1}]$. Also, when the modulation wave number is slightly smaller than $k_m$ or the length scale of excitation of harmonic modes is a bit larger than that of the most unstable mode, a few harmonic patterns coexist with the master mode. The amplitudes and widths of the solitary patterns change with time. However, the collisions and fusions among them can not occur frequently, and the spatial partial coherence of the patterns may be there, leading to the coexistence of TC and SPC states. Although the energy transfer occurs from initially excited multiple modes to a single or couple of higher harmonic modes with stronger amplitudes but reduced wavelengths, the transfer rate is not fast enough to cause STC. We have verified the existence of these ST and SPC states by analyzing the largest Lyapunov exponent spectra over the entire spatial domain and measures of the correlation function and the mutual information function over the distance (and as it increases) between any two points in space.    However, as we further reduce $k$ from a starting value of $k$ $(<k_m)$ at which the TC and SPC states coexist, there appear to be the excitation of many solitary patterns, which, as time goes on, interact themselves and fuse to form new incoherent patterns and undergo through strong electron-acoustic wave emission. The wave energy transfer occurs from initially excited many coherent modes at large scales to a single or a couple of stable higher harmonic incoherent modes with shorter wavelengths. Such a process develops faster as one approaches moderate to ultra-relativistic degeneracy regimes. The largest Lyapunov exponents are found positive in all these degenerate regimes and the vanishing of the correlations and mutual information functions over larger spatial domains.
\par {An important finding of this work is that, in ultra-relativistic degenerate plasmas $(R_0\gg1)$, the modulational instability growth rate does not always saturate within the domain $0<k<1$ (Note that since $k$ is normalized by the inverse of the plasma skin depth, the values with $k>$ are inadmissible, otherwise, the plasma collective behaviors may disappear) into stable solitons as suggested in earlier studies \cite{shatashvili2020nonlinear}, but may evolve into incoherent spatiotemporal chaos by the influence of electron-acoustic wave emission, thereby establishing a new dynamical regime beyond the soliton regime.}
\par {We mention that the spectral broadening and rapid variability in EM wave emissions from compact astrophysical objects are typically related to nonlinear wave-particle interactions. These phenomena can be associated with chaotic states and broadband power spectra. Nonlinear wave-particle interactions can cause frequency broadening and chaotic variations in radiation spectra and their characteristics. Although nonlinear wave-wave interactions can also contribute to spectral changes, especially rapid variability, these are usually the result of the superposition of different waves and may not be related to the interactions studied here.     
 Another important phenomenon, EM pulse fragmentation, is also primarily related to wave-particle interactions. In compact objects, extreme conditions lead to fragmentation of matter, which is influenced by EM waves or particles. Processes like scattering and absorption, where energy transfers between waves and particles, can drive fragmentation.   } 
\par 
To conclude, depending on the modulational perturbation wavelength smaller or larger than that of the most unstable wave, the electromagnetic radiation spectra emanating from compact astrophysical objects and interacting with electron-acoustic perturbations sustained by multi-component degenerate plasmas that surround the compact objects may either settle into EM envelope solitons (or stable wave states) due to saturation of the modulational instability,  {especially in the regime of weak relativistic degeneracy}  or undergo through strong electron-acoustic wave emissions causing to emerge spatiotemporal chaotic states,  {especially in the regime of ultra-relativistic degeneracy}, which may be signature of turbulence in dense astrophysical regimes \cite{banerjee2010spatiotemporal,misra2009pattern,chian1999}. 
{Zakharov-like models are limited to studying the generation of envelope solitons through MI, wave collapse, and wave turbulence through wave-wave interaction. However, phenomena such as the generation of dromion solutions or vortex formation, as well as those related to wave-particle interactions, cannot be studied using Zakharov-like equations.
Building on these limitations, our results show that STC can arise as a natural state in relativistic degenerate plasmas. These chaotic states may explain the origin of turbulence and fluctuations in radiation in dense astrophysical environments. Future studies with two- or three-dimensional models and comparisons with observational data will be essential to confirm these results and understand their relevance in real astrophysical systems.} 
\section*{Acknowledgments}  
One of us, SDA wishes to thank University Grants Commission (UGC), Government of India, for support through a junior research fellowship (JRF) with NTA reference no. 211610078837.   
\section*{Author declarations}
\subsection*{Declaration of Competing Interest}
The authors declare that they have no known competing financial interests or personal relationships that could have appeared to influence the work reported in this paper.
\subsection*{CRediT authorship contribution statement}
Sukhendu Das Adhikary: Formal analysis (equal); Investigation (equal); Methodology (equal); Software (equal); Writing--original draft (equal).   Amar P. Misra: Conceptualization (equal); Formal analysis (equal); Investigation (equal); Methodology (equal); Software (equal); Supervision; Validation; Writing--review \& editing.
%
\section*{Data availability statement}
The data that support the findings of this study are available from the corresponding author upon reasonable request.
\appendix
{
	\section{Pseudocode to reproduce numerical integration}\label{apx-pseudocode}
	We recast Eqs. \eqref{eq-nor EM wave} and \eqref{eq-nor EAW} as
	\begin{equation}\label{eq-F1}
		\begin{split}
			\frac{\partial A}{\partial t}=-v_{g}\frac{\partial A}{\partial x}+\frac{i}{2}\left(B_{0}\frac{\partial^2 A}{\partial x^2}+B_{1}NA
			+B_{2}|A|^2A\right)\equiv F_{1},
		\end{split}
	\end{equation}
	\begin{equation}\label{eq-F2}
		\frac{\partial N}{\partial t}=U\equiv F_{2},
	\end{equation}
	\begin{equation}\label{eq-F3}
		\frac{\partial U}{\partial t}=\frac{\partial^2 N}{\partial x^2}-3b_{3}\frac{\partial^2|A|^2}{\partial x^2}\equiv F_{3}.
	\end{equation} 
	In our numerical algorithm, we discretize time  as 
	$t = t_k = k \, \Delta t, \quad k = 0, \ldots, N_t,$  
	and space as 
	$x = x_j = j \, \Delta x, \quad j = 0, \ldots, N_x - 1.$  
	We denote the solution as 
	$u(x_j, t_k) \equiv u^k_j$,  
	and use the periodic boundary conditions: 
	$A(L, t) = A(0, t),$ $N(L, t) = N(0, t),$ and $U(L, t) = U(0, t),$    
	or, for the discretized solution, 
	$A^k_{N_x-1} = A^k_0$, $N^k_{N_x-1} = N^k_0$,  and $U^k_{N_x-1} = U^k_0$.
	The functions $F_{1}$, $F_{2}$ and $F_{3}$ contain discretized variables where we use the central difference approximations for the spatial derivatives, i.e.,
	\begin{equation*}
		\frac{\partial A}{\partial x}\approx \frac{A_{j+1}-A_{j-1}}{2\Delta x}, \quad \frac{\partial^2 A}{\partial x^2}\approx\frac{A_{j+1}-2A_{j}+A_{j-1}}{\Delta x^2},
	\end{equation*}
	for $j=1,2,\cdots N_{x}-2$. At the boundaries ($j=0$ and $j=N_{x}-1$) the spatial derivatives are defined as
	\begin{equation*}
		\left.\frac{\partial A}{\partial x}\right|_{x_{0}}\approx \frac{A_{1}-A_{N_{x}-1}}{2\Delta x}, \quad \left.\frac{\partial^2 A}{\partial x^2}\right|_{x_{0}}\approx\frac{A_{1}-2A_{0}+A_{N_{x}-1}}{\Delta x^2},
	\end{equation*}
	\begin{equation*}
		\left.\frac{\partial A}{\partial x}\right|_{x_{N_x-1}}\approx \frac{A_{0}-A_{N_{x}-2}}{2\Delta x},
	\end{equation*}
	\begin{equation*}
		\left.\frac{\partial^2 A}{\partial x^2}\right|_{x_{N_x-1}}\approx\frac{A_{0}-2A_{N_{x}-1}+A_{N_{x}-2}}{\Delta x^2},
	\end{equation*}
	where $\Delta x$ is the spatial step, and similar expressions for $N$ and $U$. }
	\par 
	{Next, we define the triplet, $\left(F_{1}, F_{2}, F_{3}\right)$ as the vector-valued function $\mathbf{F}\left(A, N, U, t\right)$, i.e.,
	\begin{equation}
		\left(F_{1},F_{2},F_{3}\right)=\mathbf{F}\left(A,N,U,t\right).
	\end{equation}
	Then the fourth-order Runge--Kutta scheme reads
	\begin{align*}
		(k_{1A}, k_{1N}, k_{1U}) &\leftarrow 
		\mathbf{F}(A, N, U, t), \\[6pt]
		(k_{2A}, k_{2N}, k_{2U}) &\leftarrow 
		\mathbf{F}\!\Big(
		A + \tfrac{\Delta t}{2} k_{1A}, \;
		N + \tfrac{\Delta t}{2} k_{1N},  \\[-2pt]
		& \hspace{1.7cm}
		U + \tfrac{\Delta t}{2} k_{1U}, \;
		t + \tfrac{\Delta t}{2}
		\Big), \\[6pt]
		(k_{3A}, k_{3N}, k_{3U}) &\leftarrow 
		\mathbf{F}\!\Big(
		A + \tfrac{\Delta t}{2} k_{2A}, \;
		N + \tfrac{\Delta t}{2} k_{2N},  \\[-2pt]
		& \hspace{1.7cm}
		U + \tfrac{\Delta t}{2} k_{2U}, \;
		t + \tfrac{\Delta t}{2}
		\Big), \\[6pt]
		(k_{4A}, k_{4N}, k_{4U}) &\leftarrow 
		\mathbf{F}\!\Big(
		A + \Delta t \, k_{3A}, \;
		N + \Delta t \, k_{3N},  \\[-2pt]
		& \hspace{1.7cm}
		U + \Delta t \, k_{3U}, \;
		t + \Delta t
		\Big), \\[6pt]
	\end{align*}
	Thus, we obtain the solutions for $A$, $N$, and $U$ at time $t+\Delta t$  as
	\begin{align}
		A &\leftarrow A 
		+ \tfrac{\Delta t}{6}\left(
		k_{1A} + 2k_{2A} + 2k_{3A} + k_{4A}
		\right), \\
		N &\leftarrow N 
		+ \tfrac{\Delta t}{6}\left(
		k_{1N} + 2k_{2N} + 2k_{3N} + k_{4N}
		\right), \\
		U &\leftarrow U 
		+ \tfrac{\Delta t}{6}\left(
		k_{1U} + 2k_{2U} + 2k_{3U} + k_{4U}
		\right).
	\end{align}
	}
	\par
	{ Typically, an explicit time integration scheme exhibits stability when the time step,
	$\Delta t$ is chosen sufficiently small such that the condition, $\Delta t < C\,(\Delta x)^{p}$ holds for some positive constant $C$. Here, the exponent, $p~(>0)$ corresponds to the order of the highest-order spatial derivative in the evolution equation \cite{Cross_Greenside_2009}. For the present model, $p = 2$ and $C ={\min}_j\left\lbrace 1/2D_j\right\rbrace$ with $D_j$ denoting the moduli of the coefficients of the highest derivatives.
	In Eqs. \eqref{eq-F1}--\eqref{eq-F3}, there are three second-order spatial derivatives, namely, $\partial^2 A/\partial x^2$, $\partial^2 N/\partial x^2$, and $\partial^2|A|^2/\partial x^2$ with coefficients $iB_0/2$, $1$, and $-3b_3$ respectively. Thus, the stability condition reduces to
	\begin{equation}
	\begin{split}
		\Delta t <& \min\left\{\frac{\Delta x^{2}}{B_{0}},\frac{\Delta x^{2}}{2},\frac{\Delta x^{2}}{6b_3}\right\}\\
		&=\Delta x^{2}\min\left\{\frac{1}{B_{0}},\frac{1}{2},\frac{1}{6b_3}\right\}\equiv\tau_{\min}.
		\end{split}
	\end{equation}
	In our simulation, we chose the time step as 
	\begin{equation}\label{eq-condition}
		\Delta t<\min\left\{10^{-4}, \tau_{\min} \right\}.
	\end{equation}
	Also, since we considered the simulation box size as $L_x=2\pi/k$ and the number of grid points as $2048$, the spatial step size is $\Delta x=L_x/2048$.	
}
\begin{table}
	{
		\begin{tabular}{|c|c|c|c|}
			\hline
			$R_0$     & $k$    & $\Delta x$ & $\Delta t$\\ \hline
			\multirow{3}{*}{$5$} & $0.15$ & $0.0205 $& $2\times 10^{-5}$ \\ \cline{2-4}
			& $0.11$ &  $0.0279$ & $3\times 10^{-5}$  \\ \cline{2-4}
			& $0.043$ & $0.0713$ & $ 10^{-4}$ \\ \hline
			\multirow{3}{*}{$25$} & $0.21$ & $0.0146$& $5\times 10^{-5}$ \\ \cline{2-4}
			& $0.11$ &  $0.0279$ & $10^{-4}$  \\ \cline{2-4}
			& $0.047$ & $0.0653$ & $ 10^{-4}$ \\ \hline
			\multirow{3}{*}{$85$} & $0.29$ & $0.0106$& $5\times 10^{-5}$ \\ \cline{2-4}
			& $0.11$ &  $0.0279$ & $10^{-4}$  \\ \cline{2-4}
			& $0.047$ & $0.0653$ & $ 10^{-4}$ \\ \hline
		\end{tabular}
		\caption{Spatial $(\Delta x)$ and time $(\Delta t)$ steps for each simulation using Eqs. \eqref{eq-condition}.}
	}
\end{table}
\bibliographystyle{elsarticle-num} 
\bibliography{Reference}
\nopagebreak
\end{document}